\documentclass[%
 aip,
 amsmath,amssymb,
 10pt,twocolumn,
]{revtex4-1}

\usepackage{graphicx}
\usepackage{dcolumn}
\usepackage{bm}

\usepackage[utf8]{inputenc}
\usepackage[T1]{fontenc}
\usepackage{mathptmx}
\usepackage{color}
\usepackage{chemfig}
\usepackage{textcomp}
\newcommand*\aap{A\&A}

\renewcommand{\H}{\mbox{H}} 
\renewcommand{\O}{\mbox{O}} 
\newcommand{\C}{\mbox{C}} 
\begin{document}

\title{Neural Network Potential Energy Surface for the low temperature
  Ring Polymer Molecular Dynamics of the H$_2$CO + OH reaction\\}


\author{Pablo del Mazo-Sevillano}
\email{pablo.delmazo@uam.es}
\affiliation{Unidad Asociada UAM-CSIC,
                       Departamento de Qu{\'\i}mica F{\'\i}sica Aplicada, Facultad de
                      Ciencias M-14, Universidad Aut\'onoma de Madrid, 28049, Madrid, Spain}
\author{Alfredo Aguado}
\affiliation{Unidad Asociada UAM-CSIC,
                       Departamento de Qu{\'\i}mica F{\'\i}sica Aplicada, Facultad de
                      Ciencias M-14, Universidad Aut\'onoma de Madrid, 28049, Madrid, Spain}

\author{Octavio Roncero}
\email{octavio.roncero@csic.es}
\affiliation{Instituto de F{\'\i}sica Fundamental (IFF-CSIC),  
                          C.S.I.C., 
                       Serrano 123, 28006 Madrid, Spain}



\begin{abstract}
  A new potential energy surface (PES) and dynamical study are presented of the reactive process
  between H$_2$CO + OH towards the formation of HCO + H$_2$O and HCOOH + H.
  In this work a source of spurious long range interactions in  
  symmetry adapted neural network (NN)  schemes is identified,
  what prevents their direct application for low temperature dynamical studies.
 For this reason, a partition of the PES into a diabatic matrix plus a NN many body term has been used,
 fitted with a novel artificial neural networks scheme that prevents spurious asymptotic interactions.
 Quasi-classical trajectory and ring polymer molecular dynamics (RPMD)  studies have been carried on this PES
 to evaluate the rate constant temperature dependence for the 
 different reactive processes, showing a good agreement with the available experimental data. 
 Of special interest is the analysis of the previously identified trapping mechanism 
 in the RPMD study, which can be attributed to spurious resonances associated
 to excitations of the normal modes of the ring polymer.
 
 \vspace*{1cm}
 \begin{center}
     Accepted in J. of Chem. Phys. (2021)
 \end{center}
\end{abstract}

\maketitle

\section{Introduction}

The raise of the reaction rate constant at low temperatures measured in different laboratories
for several organic molecules (OM) with OH and other reactants
\cite{Shannon-etal:13,Gomez-Martin-etal:14,Caravan-etal:15,Antinolo-etal:16,Ocana-etal:17,Heard:18,Ocana-etal:19}
opened a possible solution for the enigma
of how organic molecules are generated in cold dense molecular clouds, below 10 K,
in the interstellar medium. These reactions usually present
reaction barriers, which make impossible the reaction at low temperatures in gas phase. For that
reason, it was assumed that these molecules were formed on cosmic ices, in cold molecular clouds
at about 10 K \cite{Hudson-Moore:99,Watanabe-etal:07}, and then released to gas phase
as they evolve to hotter structures\cite{Garrod-Herbst:06}, such as hot cores and corinos,
where they were observed \cite{Snyder-etal:69,Ball-etal:70,Gottlieb-etal:73,Godfrey-etal:73}.
However, recently several observations of OMs were made in UV-shielded cold cores at 10 K
\cite{Remijan-etal:06,Bacmann-etal:12,Cernicharo-etal:12,Vastel-etal:14}, what
introduces the need of changing the model. Several hypotheses were introduced, such as the possibility
of chemidesorption\cite{Cazaux-etal:16,Oba-etal:18},
the incidence of cosmic rays \cite{Mainitz-etal:16} and of secondary UV photons, that may
induce the photodesorption. However, OMs usually present dipoles and strong binding energies,
and recent experiments performed by illuminating cosmic ice precursors showed that only their photofragments
are desorbed into gas phase\cite{Cruz-Diaz-etal:16,Bertin-etal:16}. This may suggest that the origin of OMs in gas phase should be
a combined mechanism, in which they are formed on ices, then their photofragments released to gas phase, where they suffer
a reprocessing to finally end in the formation of the OM.

The reprocessing could be accelerated in gas phase by the huge increase of the reaction rate observed in the laboratory below 100 K
\cite{Shannon-etal:13,Gomez-Martin-etal:14,Caravan-etal:15,Antinolo-etal:16,Ocana-etal:17,Heard:18,Ocana-etal:19}.
This increase is explained by the formation of a complex between the reactants, from where they tunnel to form
the products\cite{Ocana-etal:17,Heard:18}. However, several transition state theory (TST)
studies of the  CH$_3$OH + OH reaction \cite{Siebrand-etal:16,Gao-etal:18,Ocana-etal:19,Nguyen-etal:19}
did not get a so steep raise of the rate constant at low temperature, but it was much slower than the experimental results.
They all attributed
the difference to pressure effects, which at microscopic level may be attributed to the formation of complexes,
between the reactants and of them with the buffer gas used in the expansions. These complexes could experience
secondary collisions adding the extra energy needed to overcome the barrier. However, due to the low densities of molecular clouds
 the zero-pressure rate constant is needed in their models. It is therefore mandatory
to check if TST methods can describe quantitatively the tunneling in the deep regime.

In this line, we have studied the dynamics of the reactions of OH with H$_2$CO and CH$_3$OH
\cite{Zanchet-etal:18,Roncero-etal:18, delMazo-Sevillano-etal:19, Naumkin-etal:19}.
For this purpose, full dimensional potential energy surfaces (PESs) were fitted to accurate {\it ab initio} calculations
\cite{Zanchet-etal:18,Roncero-etal:18}.
For H$_2$CO + OH reaction the QCT rate constant yield a semiquantitative agreement with experimental results  \cite{Ocana-etal:17},
while for CH$_3$OH + OH the simulated rate constant was far too low \cite{Roncero-etal:18}. To check if quantum
effects could explain the difference, ring polymer molecular dynamics (RPMD) calculations were performed on the two PESs
\cite{delMazo-Sevillano-etal:19},
finding an excellent agreement in the case of methanol, but a similar semiquantitative agreement for the case of H$_2$CO.
These good agreement could explain the experimental results and provide a good estimate of the
zero-pressure rate constant at low temperature,
but several problems still persist. First, below 100 K RPMD leads to a high trapping probability, 
forming complexes with extremely long lifetimes (larger than several hundreds of nanoseconds), whose trajectories cannot
be finished. Therefore, the reaction rate has to be evaluated by multiplying the trapping rate constant by a ratio to products
that is inferred from TST method or QCT calculations \cite{delMazo-Sevillano-etal:19}. This could also be due to the PES,
and the second problem is to improve the accuracy
of the fit. This is the goal of this paper, by applying Artificial Neural Networks (ANN) techniques to fit a new PES to
check if the high trapping rate persist, in one side, and to improve the comparison with experimental data.

In this work we shall explore different alternatives for multidimensional fitting with ANN techniques,
including the symmetry either with permutationally invariant polynomials,  PIP-NN\cite{Jiang2013, Li:13, Li2015, Jiang:16, Jiang:2020}, 
or with fundamental invariants,  FI-NN\cite{Shao2016,Lu2018}. In addition, the channel towards
HCOOH+H products, absent in the previous PES\cite{Zanchet-etal:18}, will be included.
We shall focus the attention in the problem
of including long-range interactions, which are crucial to simulate the dynamics at low
temperature. We shall then use the new PES to perform QCT and RPMD calculations to compare with the
experimental results and to analyze the formation of long-lived complexes.

This paper is organized as follows. Firstly, a description of {\it ab initio} calculations
is provided in section 2. A general description of artificial neural networks is then provided in section 3,
focusing on the  description of 
a new many-body scheme to accurately include the long-range interactions
crucial in the dynamics at low temperature in section 4. With this, an overall
description of the PES will be provided in section 5, showing the most important
features. In section 6,  we will focus on the dynamical results
on this new PES, employing both QCT and RPMD methodologies, comparing with previous theoretical and
experimental results. Finally some conclusions
will be extracted together with several crucial questions to be addressed in the future.

\section{\textit{Ab initio} calculations}

The reactions
\begin{subequations}
  \begin{eqnarray}
	\H_2\O\H + \O\H(^2\Pi) &\rightarrow&  \H\C\O + \H_2\O \label{eqn:reac1}\\
	&\rightarrow&  \H\C\O\O\H + \H \label{eqn:reac2}
\end{eqnarray}
\end{subequations}
are studied using the explicitly correlated coupled cluster method, RCCSD(T)-F12a/cc-pVDZ\cite{doi:10.1063/1.3054300},
as implemented
in the MOLPRO package\cite{MOLPRO_brief}. 
CCSD(T) calculations are considered a benchmark, and
it has recently been shown that 
CCSD(T)-F12 is able to reach results close to CCSD(T)/CBS, even with
relatively small basis sets \cite{Adler2007, C3CP44708A, doi:10.1063/1.3054300}. 
The effect over reactivity from an excited electronic state has been 
despise from the results of MRCI/cc-pVDZ calculations along the 
RCCSD(T)-F12 minimum energy path, since it becomes highly repulsive
as the systems approach the transition states of either HCO + H$_2$O
or HCOOH + H reaction channels. 
The reader is referred to the Supplementary Information for a further description.

The calculated stationary points at this level of theory are rather similar to those previously reported
\cite{Zanchet-etal:18, Francisco1992, Alvarez-Idaboy2001,Barbara2003,AkbarAli2015,Wang2013,DeSouzaMachado2020}.
 The energy of the transition state to form HCO + H$_2$O is very low (of $\approx$ 27 meV in this work) 
and is the point where larger discrepancies are found among the 
different works, depending on the level of theory adopted. This low energy
is of the order of the expected accuracy for a system with so many electrons,
and the calculated height varies from positive to negative depending of the level of theory chosen.
The calculations presented here are considered to be highly accurate for the 
available methods at the moment, keeping the calculation of single points rather
affordable for the calculation of the full dimensional PES.

Reaction \ref{eqn:reac1} shows a single transition state (TS)  to form HCO + H$_2$O, called
TS1 (see figure \ref{fig:MEP}), with an energy of $27.1$ meV. Before this transition state
a minimum (RC1) is found at an energy of $-235.8$ meV.
The minimum energy path (MEP) to form HCOOH + H, in Eq.~\ref{eqn:reac2}, is more complex.
The system must
surpass the transition state (TS2), where the OH lies over the H$_2$CO
plane. Then, the H$_2$CO bends and, at the same time, the OH gets closer to
the carbon atom. Finally, one of the H$_2$CO hydrogens leaves, forming
HCOOH. The formation of $c-$HCOOH or $t-$HCOOH depends on whether the
OH rotates or not from the stationary point RC4.

For the present fit, around 180000 {\it ab initio} points were added
to the previously calculated for the  PES of Ref.~\cite{Zanchet-etal:18}.
This was needed to increase the accuracy of reaction \ref{eqn:reac1}, already explored
in Ref.~\cite{Zanchet-etal:18}, and to describe, for the first time, the channel
towards HCOOH + H, in reaction \ref{eqn:reac2}. The new points were calculated iteratively including
points for the minimum energy paths (MEPs) and normal modes along them
evaluated on intermediate fits. Also QCT and RPMD trajectories were
ran on intermediate versions of the fits to populate physically accessible configurations.
For each new point, the Euclidean distance \cite{Jiang2013} to all the 
other points is evaluated as $d_{ab}=\sqrt{\sum_{i=1}^{15} (d_i ^a - d_i^b})^2$, 
where $d_i^a$ is the $i$th interatomic distance of geometry $a$.

\section{Artificial Neural Networks}

The use of Permutationally Invariant Polynomials (PIP) with Rydberg-type functions 
has been widely used to describe the PES of polyatomic systems with no many atoms\cite{Aguado-Paniagua:92,Braams-Bowman:09}.
In these methods, few non-linear exponential parameters are used, while the linear coefficients of
each term are fitted with a least square method. 
In order to obtain a PES that is invariant with respect to translation, rotation and inversion,
the fit of the potential is carried out as a polynomial in terms of the internuclear
distances\cite{Aguado-etal:94}.
As the number of internuclear distances increases more than the number of independent coordinate
it means that, even considering low orders of the fitted polynomial, the number of coefficients
necessary to carry out the fit increases enormously with a small increase of the number of atoms
in the system. This implies a limitation in the application of these methods to systems of more
than 6-8 atoms.
Instead, in this work we make use of neural networks (NN) to fit a new PES.

Feed forward neural networks\cite{MURTAGH1991183} (FFNN) are a type of functions able to transform a vectorial
input into a vectorial output, in contrast to other architectures devoted to work
on matrix\cite{ciresan2011flexible} or graph\cite{duvenaud2015convolutional} representations.
This kind of neural networks has been of great
interest in the field of PES fitting, where a molecular representation
of the system is given to the neural network, returning its potential energy. This 
algorithm works in a supervised way, since the neural network is trained 
against a set of target energies, \textit{e.g. ab initio}, usually minimizing the 
root mean squared error (RMSE) with the predictions.

\begin{figure}[hbtp]
  \centering
  \includegraphics[width=0.7\linewidth]{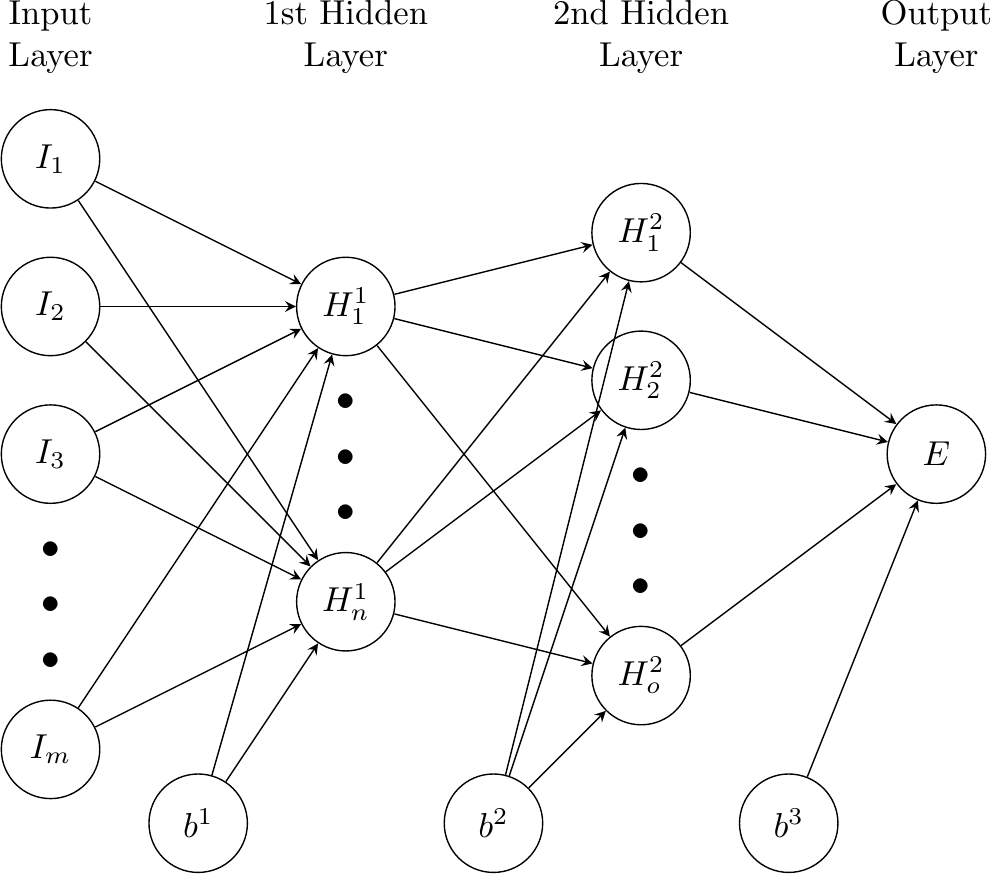}
	\caption{Computational graph of a FFNN with two hidden layers.}
  \label{fig:NN_graph}
\end{figure}

A FFNN is characterized by the number of layers  and 
the number of neurons at each layer, increasing the flexibility of the 
function as more layers and neurons are included. The computational graph for such a FFNN with 
two hidden layers is presented in figure \ref{fig:NN_graph}. A simple rule is given to move forward in the
layer structure, from the input (molecular geometry) to output (energy):
\begin{equation}
  H_j^l = \sigma \left( \sum_i w_{ji}^{l} H_i^{l-1} + b_j^l \right),
  \label{eqn:forward_move}
\end{equation}
where $H_j^l$ is the $j$th value of the $l$th hidden layer, $w_{ij}^l$ and $b_j^l$ are the
trainable weights and bias in that layer and $\sigma$ is the transfer function which is,
in general, not linear.

Typical choices of $\sigma$ are hyperbolic tangent or sigmoid functions between the 
hidden layers in the context of PES fitting ,
and a linear function between the last hidden layer and output layer to map the whole range of real numbers.
  In this work we used a logistic function, $\sigma(x) = 1/(1+\exp(-x))$,
between the hidden layers. The input representation
for each ANN is defined as the negative exponential of each interatomic distances. The corresponding PIP of FI
polynomials are evaluated on them whenever the permutational symmetry is considered. Finally, all input values 
are standardize to zero mean and unit standard deviation. The PIP and FI functions used are presented in the 
Supplementary Information.

The NN structure will be expressed in the usual way, I-H$_1$-$\cdots$-H$_n$-O,
with I and O the number of neurons in the input and output layers, and $n$
hidden layers, with H$_i$ neurons in the $i$th hidden layer.
PIP-NN and FI-NN  schemes, including
the permutational symmetry of identical atoms, have
many properties in common so artificial neural networks (ANN) will be used to refer indistinctly to 
either PIP-NN or FI-NN in this paper.

For each permutational symmetry group, the FI are calculated with the King's algorithm
\cite{King2013} implemented in the computer algebra system \textit{Singular} \cite{DGPS}. For many
systems these FI have already been evaluated and FORTRAN routines for their evaluation are provided
in \url{https://github.com/pablomazo/FI}, expanded from the original repository \url{https://github.com/kjshao/FI}.

For training ANNs we have developed NeuralPES, a Python code based on PyTorch\cite{NEURIPS2019_9015} library,
meant to assist at each step of the training process, beginning with data preprocessing, hyperparameter
tuning and training itself.

\section{\label{sec:longrange} Long range interaction with neural networks}

The long range behaviour of a PIP PES was analyzed in Ref.\cite{doi:10.1063/1.4811653}, concluding
that certain terms of the PIP expansion lead to spurious interactions in
the asymptotic region due to polynomials involving distances between
unconnected fragments. This was solved removing some of the polynomials
that connect the fragments asymptotically in what
is called ``Purified Basis'' \cite{Yu2016}.
 An equivalent solution 
has been recently proposed by J. Li \textit{et al.} \cite{Li2020}
to be used in the fit of many body (MB) terms with PIP-NN 
by removing those PIPs relating unconnected distances.

In this paper we demonstrate that, on top of these terms
of unconnected distances, there exist another source of spurious 
interactions due to the non linearity of the transfer function in the
ANN. For doing so, a simplified A$_2$BC system is considered, focusing on the
 asymptote A$_2 \cdots $BC, such that there is no interaction between them. 
A PIP PES is defined as
\begin{equation}
  V = \sum_i c_i PIP(i)
\end{equation}
where $c_i$ are the coefficients to fit and PIP(i) is the $i$th
permutational invariant polynomial. The generator of all the
PIPs ($G$) for this particular permutational symmetry is:
\begin{equation}
  G = h_{A^1A^2}^{l_1}   h_{BC}^{l_6} \left[ h_{A^1B}^{l_2} \cdot h_{A^1C}^{l_3} \cdot h_{A^2B}^{l_4}
    \cdot h_{A^2C}^{l_5} +  h_{A^2B}^{l_2} \cdot  h_{A^2C}^{l_3} \cdot h_{A^1B}^{l_4} \cdot h_{A^1C}^{l_5} \right]
\end{equation}
where $h_{ij} = f(d_{ij})$, with $d_{ij}$ the distance between atoms
$i$ and $j$, and $l_{ij}$ is and exponent. It is common to define
$h_{ij} = \exp(-\alpha d_{ij})$, such that $h_{ij}$ tends to zero
as the distance tends to infinity. In this situation only the 
PIPs where $l_2=l_3=l_4=l_5=0$ survive in the long range:
\begin{equation}
G^{(\infty)} = h_{A^1A^2}^{l_1}   h_{BC}^{l_6}
\end{equation}
Setting $\max(l_i) = 1$ the following set of PIPs is defined:
\begin{equation}
  PIP = \{h_{A^1A^2}, \,  h_{BC}, \, h_{A^1A^2}\cdot h_{BC}\}
\end{equation}
Therefore, the PES in the asymptote is evaluated as
\begin{equation}
  V^{(\infty)} = c_1 \cdot h_{A^1A^2} + c_2 \cdot h_{BC} + c_3 \cdot h_{A^1A^2}\cdot h_{BC}
\end{equation}
Notice that, through the third term in the expansion a force between
A$_2$ and BC fragments arises, even though they were set at an infinite
distance. This PIP set could be purified by just removing the third
function, hence disconnecting both fragments. At this point, the 
asymptotic potential energy surface would depend linearly on both
fragment distances, so no spurious interactions are introduced up to now.

The next step is to show that even when the purified PIP set is used,
PIP-NN introduces spurious interactions. In a PIP-NN with only one 
hidden layer, the energy is evaluated as
\begin{eqnarray}
  V &=& \sum_j w_{Ej} H_j + b_E \\
  H_j &=& \sigma \left( \sum_i w_{ji} PIP(i) + b_j \right),
\end{eqnarray}
where $w$ and $b$ are the weight matrices and bias vectors, $\boldsymbol{H}$ 
is the hidden layer and $\sigma$ is the transfer function, which in
general is not linear. The values of the neurons of the hidden layer
evaluated on the purified PIP set become
\begin{equation}
  H_j = \sigma \left(w_{j1} \cdot h_{A^1A^2} + w_{j2} \cdot h_{BC} + b_j \right).
\end{equation}
Since $\sigma$ is not linear, a connection between both fragments 
is introduced and, with that, an spurious force between them arises.
$\partial V / \partial h_{A^1A^2}$, which ultimately
relates with the force $A^1$ and $A^2$ experience is
\begin{equation}
  \frac{\partial V}{\partial h_{A^1A^2}} = \sum_j w_{Ej} w_{j1}
  \sigma' \left(w_{j1} \cdot h_{A^1A^2} + w_{j2} \cdot h_{BC} + b_j \right),
\end{equation}
being clear that there is always a dependence with the BC fragment.

This same derivation can be develop with a set of FI, arriving to 
the same conclusion. Moreover, this problem is magnified when more
than one hidden layer is introduced. This spurious interactions lead to an energy transfer between
the fragments, that will have a critical effect in dynamical studies,
specially at low collision energies, where trajectories
could even be reflected. 
We may then conclude at this point, that
the main source of spurious interactions in a PIP-NN or FI-NN
is the non linearity of the transfer function.

Although it would yield to a very low fitting error,
the above arguments show that it is not possible to fit the PES
with one ANN since a residual interaction persists between the
reactants at long distances. This causes small changes in the internal energies
of each fragment, introducing errors that strongly affect the dynamics
at low energies, making inconvenient the use
of PIP or FI-NN PES that expand the whole configuration space.

One possible solution is to separate
the PES in two regions, short and long range, connected by a switching function \cite{Li2014a}.
The long range term would be a well behaved separable function and the short range an ANN PES.
  The main difficulty with this functional form is to tune
the switching function to produce a smooth change between both regions, 
specially as the dimensionality of the system increases. This can also have
a huge effect when studying low energy or temperature dynamics of the
system, due to the possible presence of spurious matching problems.

In this work we propose to use a partition of the potential energy
as the one used in the previous PESs\cite{Zanchet-etal:18,Roncero-etal:18,Aguado-etal:10,Sanz-Sanz-etal:13},
were a diabatic matrix, $V^{diab}$, is used to 
produce a first order PES corrected with a six-body term, as expressed by
\begin{equation}
  V = E_0^{diab} + V^{6C},
\end{equation}
where $E_0^{diab}$ is the lowest eigenvalue of $V^{diab}$. This matrix is of
dimension $N \times N$, being $N$ the number of rearrangement channels,  each of them
describing   reactans or products fragments, and the interaction between
them, including long range interactions. This matrix is analogous to those previously
used \cite{Zanchet-etal:18,Roncero-etal:18}, and is described in detail in the SI.
In the present implementation of $V^{diab}$, the accuracy of each fragment of the
PES has been increased by replacing them by ANN fittings.

The main advantage of this form is that the diabatic matrix captures
the basic features, namely the long range interactions in each rearrangement
channel and a first approximation to short range interactions,
only using PESs for each of the fragments and their interactions
on each rearrangement channel. This idea is based on triatomics-in-molecules (TRIM) \cite{Aguado-etal:10, Sanz-Sanz-etal:13}
  and diatomic-in-molecules (DIM)\cite{doi:10.1021/ja00905a002, doi:10.1021/ja00905a003}, 
  that allows an accurate description of the fragments of physical
relevance as well as the interactions among them, both short and long range.

\section{\label{sec:PES} Potential Energy Surface: NN Six-body term }

For the six-body term, $V^{6C}$, we propose to use the partition 
\begin{equation}
  V^{6C} = S \cdot  V^{ANN},
\end{equation}
where $S$ is the switching function, designed
to be zero as the system tends to any of the asymptotic regions,
removing the spurious interaction from the  $V^{ANN}$, the PIP-NN or FI-NN function,
trained to the 
difference between the \textit{ab initio} energies
and $E_0^{diab}$, $i.e.$ it should tend to zero at long distances.
In this way, the switching
function ensures once more that the MB term will go to zero in the desired regions, and the 
fragments in the asymptotic region will not be affected by spurious interactions.
On top of that, notice that it is not necessary to purify the set of invariants anymore,
since is the $S$ function which removes the spurious connections.

\begin{figure}[hbtp]
  \centering
  \includegraphics[width=0.8\linewidth]{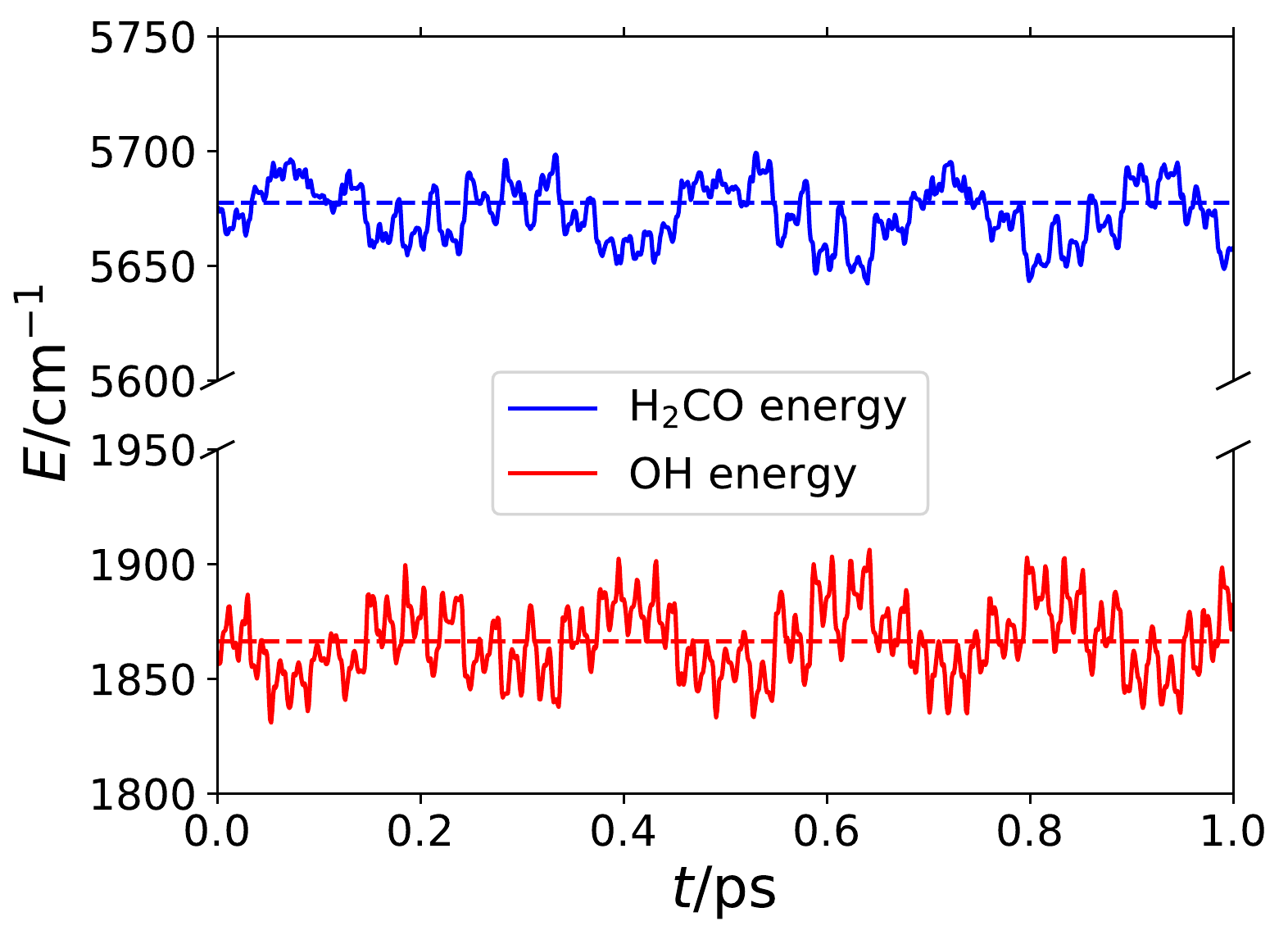}
  \caption{Energy of H$_2$CO and OH fragments in two QCT trajectories with the fragments set at a large distance
  so there is no physical interaction between them. With dashed lines a trajectory in which $V^{6C} = S \cdot V^{ANN}$
and with solid lines one where $V^{6C} = V^{ANN}$.}
  \label{fig:energy_transfer}
\end{figure}

In order to stress the importance of removing the long range spurious 
interaction in ANN PES, the energy of the H$_2$CO and OH fragments has been 
followed along two QCT trajectories, being both fragments well separated so that there is no
physical interaction between them. In the first trajectory, $V^{6C}=V^{ANN}$ and in the second one 
$V^{6C}=S \cdot V^{ANN}$. The results are shown in figure \ref{fig:energy_transfer}.
It is clear that when the $S$ function is not included to build the six-body
term, both fragments experience an spurious interaction that makes their energy change
along the trajectory. This transfer is of about 10 meV, comparable to the relative
kinetic energy at low collisional energy. This interaction is completely removed by
the switching function as it becomes evident by the constant energy of the fragments
in dashed lines.

The asymptotic channels to consider in this system are H$_2$CO + OH, 
HCO + H$_2$O and HCOOH + H. The switching function ($S$) is 
a product of hyperbolic tangents depending on
the distances $d_{CO}$, $d_{CH_1}$ and $d_{CH_2}$,
so permutational symmetry of H$_2$CO is preserved.

$V^{ANN}$ is fitted with NeuralPES program through a FI-NN with a structure 21-80-80-1, meaning
that 21 fundamental invariants are required for taking into account 
the permutational symmetry of the two formaldehyde hydrogen atoms. The FIs are
evaluated over the negative exponential of the interatomic distances. The reader
is referred to the Supplementary Information for the hyperparameter description used here.

\begin{table}[hbtp]
\centering
\caption{RMSE evaluated over each energy range for the diabatic term and the full PES.}
\begin{tabular}{c c c c}
\textbf{$\boldsymbol{E}/eV <$} & \textbf{Points} & $\boldsymbol{E_0^{diab}}$ \textbf{/ meV} & $\boldsymbol{V}$ \textbf{/ meV}\\\hline
 $0.0$    &       87356 & $369.38$ &    $ 39.18$\\
 $1.0$    &      218522 & $411.41$ &    $ 51.98$\\
 $2.0$    &      245978 & $469.64$ &    $ 65.79$\\
 $3.0$    &      261437 & $523.92$ &    $ 86.45$\\
 $4.0$    &      269743 & $564.00$ &    $ 92.05$\\
 $5.0$    &      274326 & $584.60$ &    $105.60$\\ \hline
\end{tabular}
\label{tab:pes_total_error}
\end{table}

The RMSE of $E_0^{diab}$ and the fitted PES are presented in table
\ref{tab:pes_total_error}. The ANN $V^{6C}$ term reduces the
fitting error by almost a factor of 10 for the 
lower energy ranges.

\begin{figure*}[hbtp]
	\begin{minipage}{0.5\linewidth}
		\centering
		\includegraphics[width=0.9\linewidth]{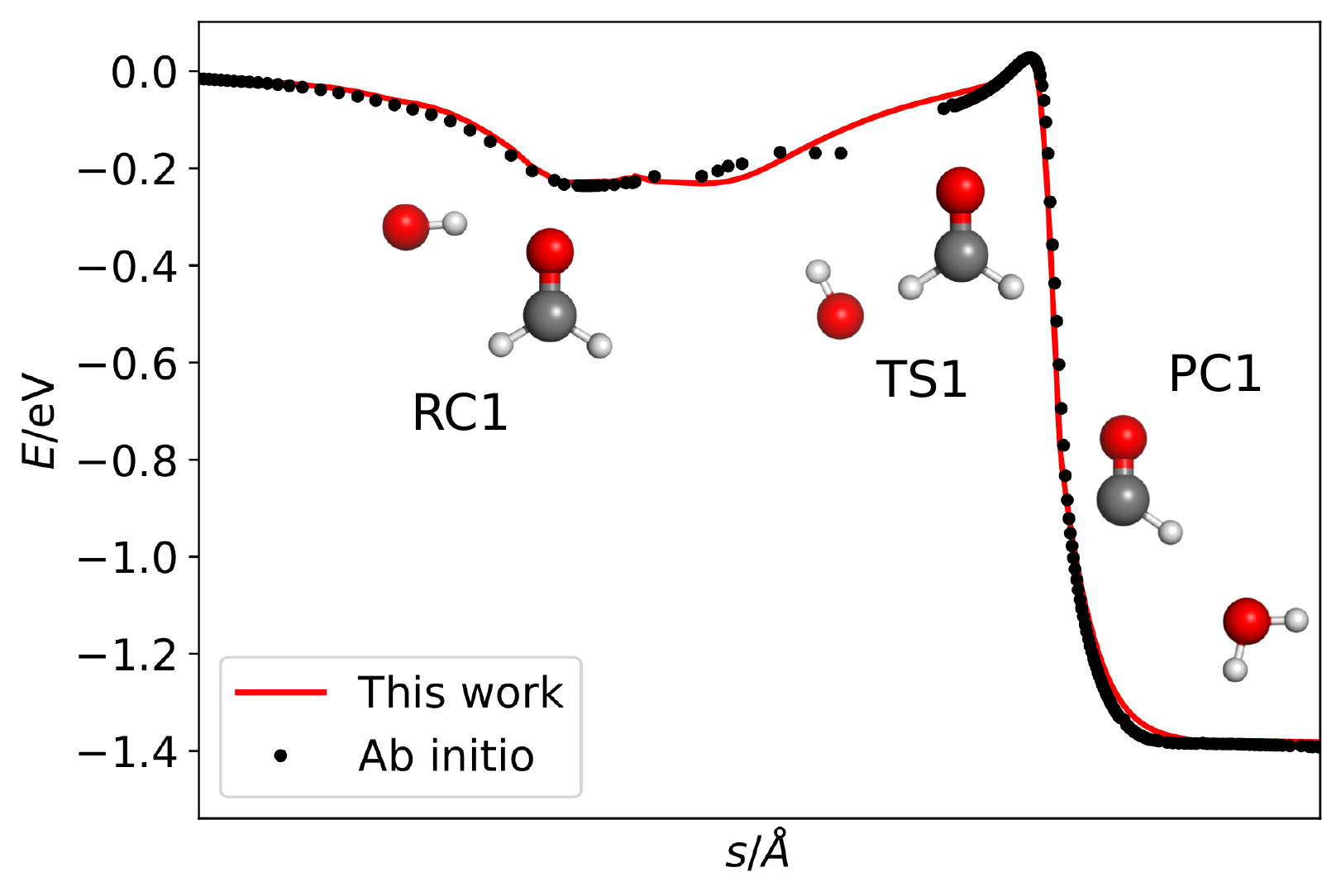}
	\end{minipage}%
	\begin{minipage}{0.5\linewidth}
		\centering
		\includegraphics[width=0.98\linewidth]{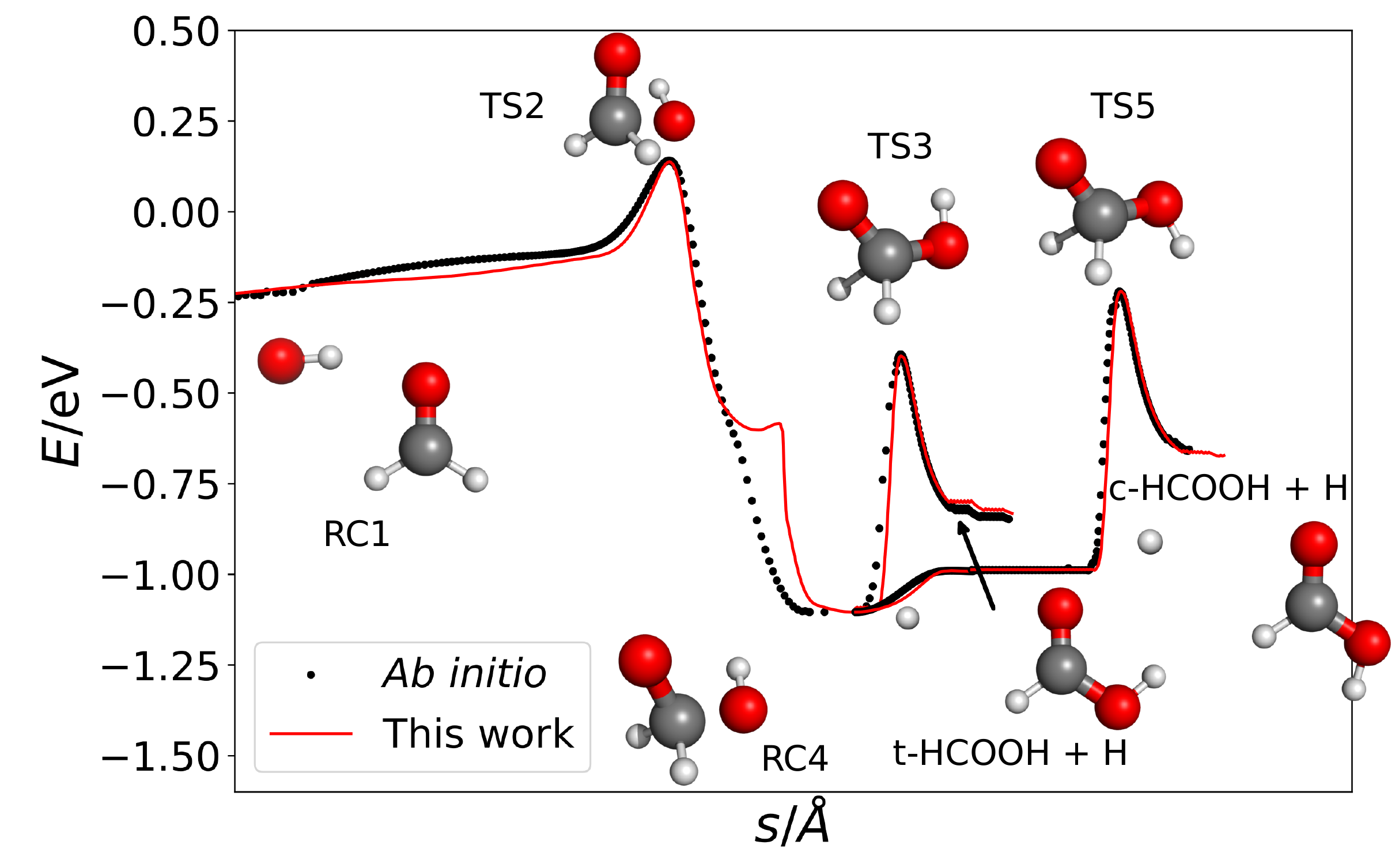}
	\end{minipage}
	\caption{Minimum energy paths for the formation of HCO + H$_2$O
	(left) and HCOOH + H (right), both \textit{cis} and \textit{trans} rotamers. 
  The geometries of the stationary points are represented along the path.}
	\label{fig:MEP}
\end{figure*}

\begin{figure}[hbtp]
  \centering
  \includegraphics[width=0.7\linewidth]{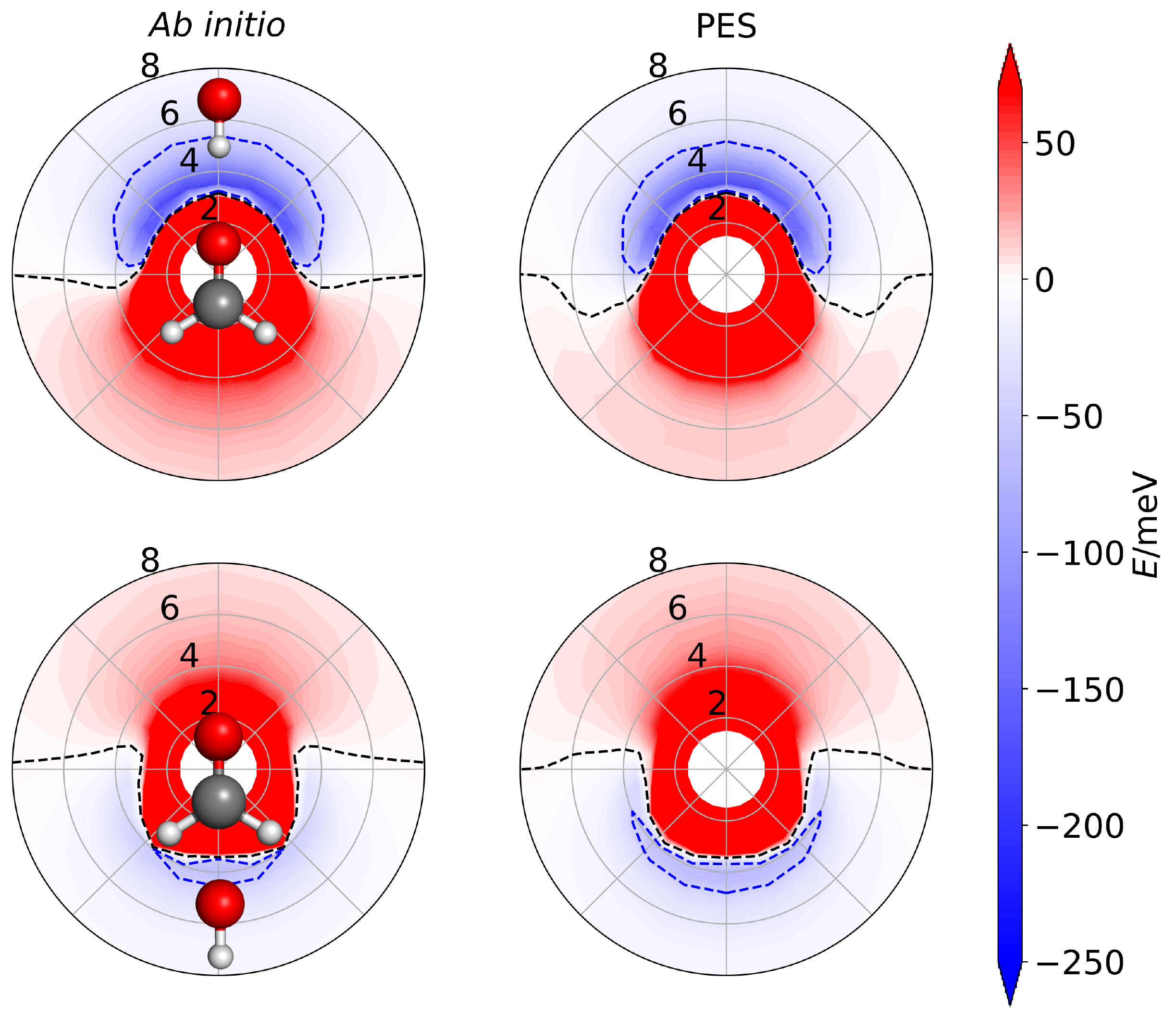}
  \caption{Energy of the system when OH approaches H$_2$CO in the plane.
	On the top (below) panels, the H (O) of OH always faces the H$_2$CO molecule. Black and blue isolines
represent energies of 0 meV and -50 meV, respectively. Distance is expressed in \AA. The H$_2$CO and OH fragments
are only meant to serve as a reference, they are not at scale.}
\label{fig:cut_short}
\end{figure}

In figure \ref{fig:MEP} the 
\textit{ab initio} and the PES minimum energy paths for reactions to
HCO + H$_2$O and HCOOH + H are shown. In both cases we find a very 
good agreement between the paths, with the only exception of a shoulder
that appears in the HCOOH + H products channel. This is not expected to have any
effect on the dynamics since it is well below the reactants asymptote.
We find an improvement with respect to the previous PES in the description
of the region of stationary point RC1, which was geometrically not well 
located. Also, another local minima on the reactants rearrangement channel, not related
to the minimum energy path, have been improved, what is of importance for a
reaction with such a roaming behaviour.

The short range region of the PES is compared against \textit{ab initio} results
in figure \ref{fig:cut_short}. In general, the PES is very well behaved, showing
a perfect agreement for both attractive and repulsive regions. 

A very good agreement is also found in the normal mode frequencies
on the different transition states as can be checked in the table
presented in the Supplementary Information. High frequency modes have
been considerably improved, in part due to a better description of the 
fragments PESs. 
It is also found that, in general, low frequency modes present greater
differences, since they relate with flatter regions of the PES that can
be more challenging to fit.

\section{Dynamical Results}

\subsection{QCT results}
\begin{figure}[hbtp]
  \centering
  \includegraphics[width=.9\linewidth]{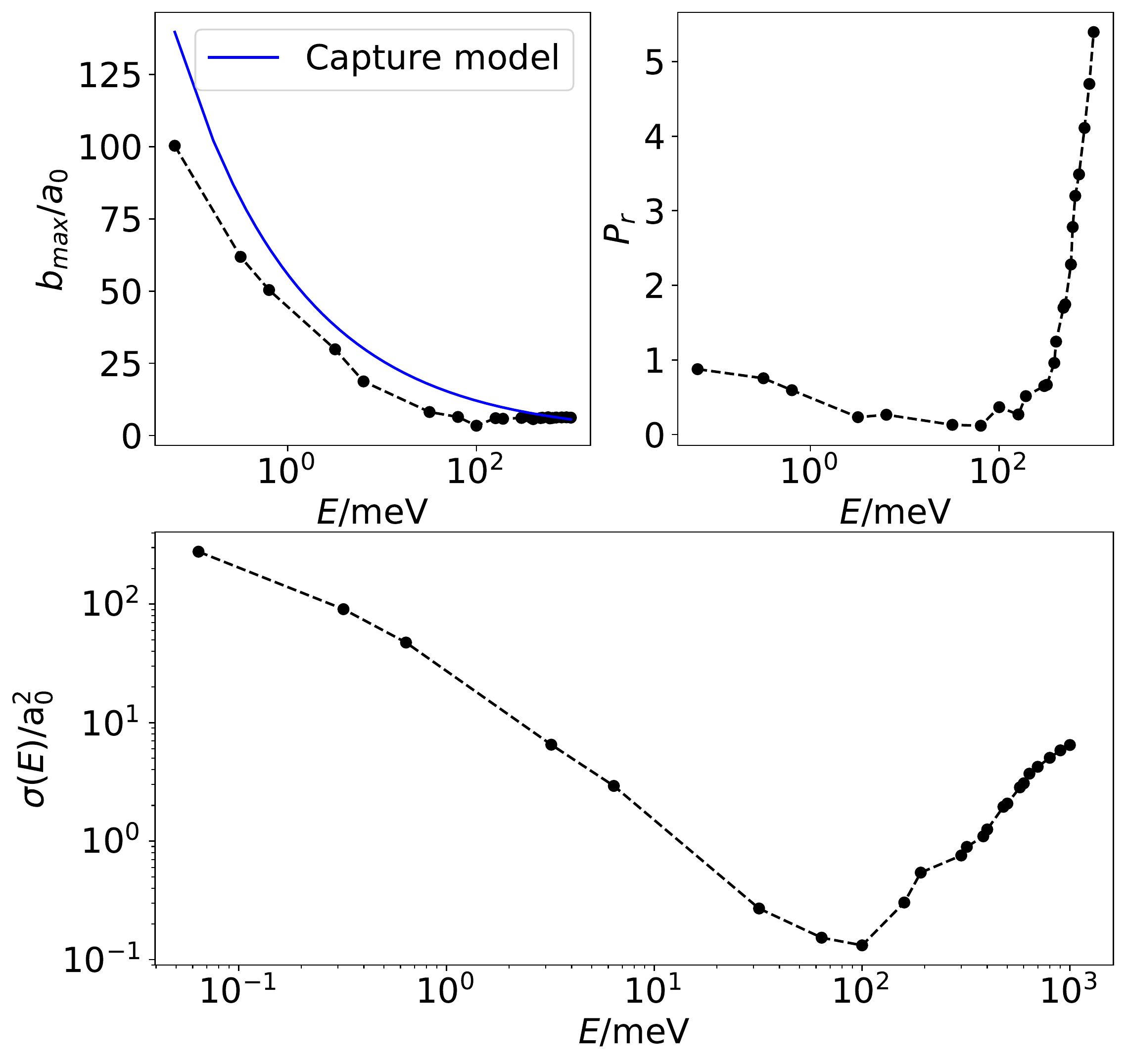}
  \caption{QCT results at fixed collisional energy for the formation of HCO + H$_2$O. On the top
  left panel, the maximum impact parameter, on the top right panel the reaction probability and in the lower panel,
the reactive cross section.}
  \label{fig:sigma_E}
\end{figure}

QCT calculations were carried with an extension of miQCT code\cite{Dorta-Urra2015, C6CP00604C}
for  $N$ atom systems. The reaction cross section $\sigma(E)$ is calculated at 
fixed collision energy with the reactants in their ground vibrational
and rotational states as:
\begin{equation}
  \sigma(E) = \pi b_{max}^2 P_r(E)
\end{equation}
where $b_{max}(E)$ is the maximum impact parameter and $P_r(E)$ the 
reaction probability at a given collision energy. 
The initial conditions of the internal degrees of freedom of the fragments have been calculated
with the adiabatic switching method \cite{Ehrenfest1917, Nagy2017}. The remaining initial conditions
are set by random numbers according to the usual
 Monte Carlo method \cite{Karplus1965}. The initial maximum impact parameter
is set according to the capture model for a dipole-dipole interaction \cite{Levine-1987, Zanchet-etal:18}
shown in Fig. \ref{fig:sigma_E} along with the converged $b_{max}$.

The reactive cross section for the formation of HCO + H$_2$O is presented
in figure \ref{fig:sigma_E}, calculated from more than $5 \times 10^{4}$
trajectories at each collisional energy within $b_{max}(E)$. The present QCT results
have the same qualitative behaviour as those
obtained with a previous PES\cite{Ocana-etal:17,Zanchet-etal:18},
and reactive cross section experiences a huge increase as the 
collisional energy decreases below 100 meV, even at energies lower than the transition
state energy barrier, at 27 meV. This increase in the reactive cross section is
explained  by an enormous increase of the maximum impact parameter,
while the reaction probability remains constant or slightly increases.

The reaction at high relative collisional energy is characterized by a
direct mechanism in which OH and H$_2$CO collide and either react or
fly apart. At low collisional energies the reaction is dominated 
by the long range, dipole-dipole, interaction which rotates and orients the reactants
 as to maximize their interaction, capturing the trajectory for
high values of the impact parameter. 
This leads to a rotational excitation of the system as the
reactants become closer that traps the system for long times, since it is
improbable to stop the rotation turning it into translation.
Direct and trapped QCT trajectories are similar to the RPMD ones, shown in figure
\ref{fig:Rx_Rz}, with the difference that trapped trajectories live much less, as will be discussed below.

The reaction probability does not got to zero for collisional energies below the potential 
barrier, since the zero-point energy (ZPE) is enough to overcome it. Still, there must exist 
couplings between the orthogonal degrees of freedom that promote the energy transfer among them. 
This coupling becomes favoured by the roaming-like mechanism and the huge complex lifetimes 
experienced at low collisional energies.


The QCT reaction rate constants have been calculated at constant temperature, with the
reactants in their ground vibrational states and considering their rotational distribution.
Only the two states that correlate with the ground spin-orbit state, OH($^2\Pi _{3/2}$), react,
so the electronic partition function is
\begin{equation}
  q_e(T) = \frac{1}{1 + \exp(-200.3/T)},
\end{equation}
and the reaction rate constant is evaluated as
\begin{equation}
  k(T) = q_e(T) \sqrt{\frac{8 k_B T}{\pi \mu}} \pi b_{max}^2(T)P_r(T)
  \label{eqn:k_T},
\end{equation}
where $b_{max}(T)$ and $P_r(T)$ are the maximum impact parameter and
reaction probability at constant temperature, and $\mu$ is the reduced
mass of the OH, H$_2$CO system.

\begin{figure}[hbtp]
  \centering
  \includegraphics[width=.9\linewidth]{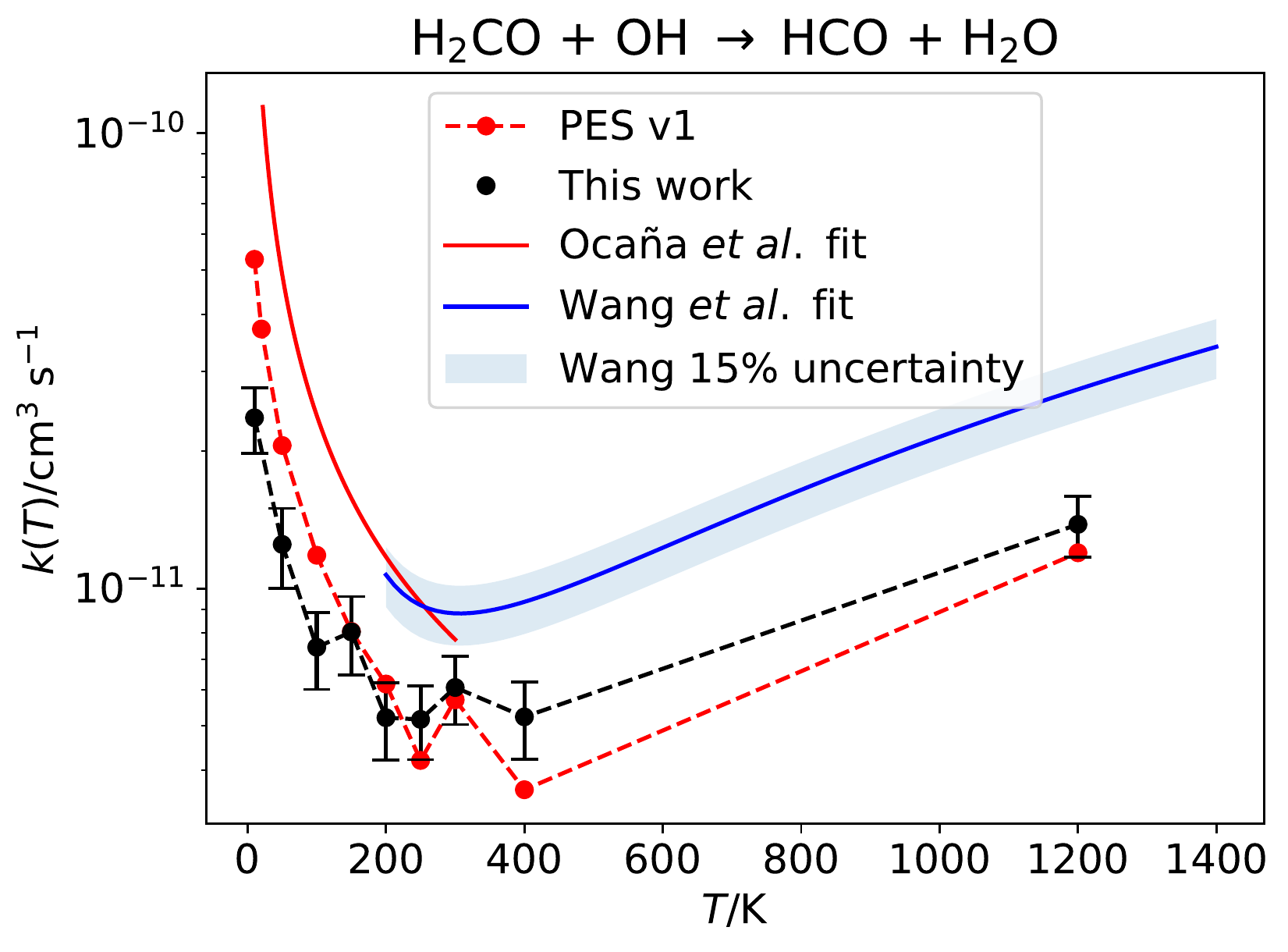}
  \caption{QCT reaction rate constant at constant temperature for HCO +
	H$_2$O formation. With black points, the results obtained in this work. 
	The error bars are two times the standard error of the Monte Carlo
	integration \cite{Karplus1965}.
	With
red points, the results from our previous PES \cite{Zanchet-etal:18}, With
solid lines, the recommended fit from Ocaña \textit{et al.} \cite{Ocana-etal:17}
in red and from Wang \textit{et al.} \cite{Wang2015} in blue.}
  \label{fig:k_QCT}
\end{figure}

The calculated reaction rate constants are shown in figure \ref{fig:k_QCT}
along with the experimental results in the temperature range from 22 K
to 1200 K, together with rates calculated earlier with another PES
\cite{Ocana-etal:17,Zanchet-etal:18}.

The reaction rate constant behaviour with temperature is basically the
same for the  two PESs above 100 K, but below this temperature there is a
quantitative difference between both of them, although both reproduce
the increase experimentally shown.
It is not surprising that the reaction rate constants calculated with the two PESs
is not well reproduced at high temperatures, since only the vibrational
ground state of the fragments is being considered. Still, there is a 
factor of about 1.5 between our results and the experimental  ones from Wang \textit{et al.}
at 300 K. This difference increases as the temperature lowers, with
respect to experimental results and also between both PES. The 
discrepancies between the two PESs could be attributed to variations in
the transition state region. Due to the relatively small
height of the TS and the high anharmonicity of this region, tiny 
differences, below the fitting error, can lead to differences in 
dynamical behaviour.

The reaction to form HCOOH + H is secondary with respect to the
formation of HCO + H$_2$O. In particular, R. A. Yetter \textit{et al.}
\cite{Yetter1989} measured the kinetic rate constants for the formation
of HCO + H$_2$O to be $(7.75 \pm 1.24) \times 10^{-12}$ cm$^3$ molecule$^{-1}$ s$^{-1}$,
in good agreement with the present results,
and $0.2^{+0.8}_{-0.2} \times 10^{-12}$ cm$^3$ molecule$^{-1}$ s$^{-1}$ 
  for HCOOH + H at 300 K.

\begin{figure}[hbtp]
  \centering
  \includegraphics[width=0.9\linewidth]{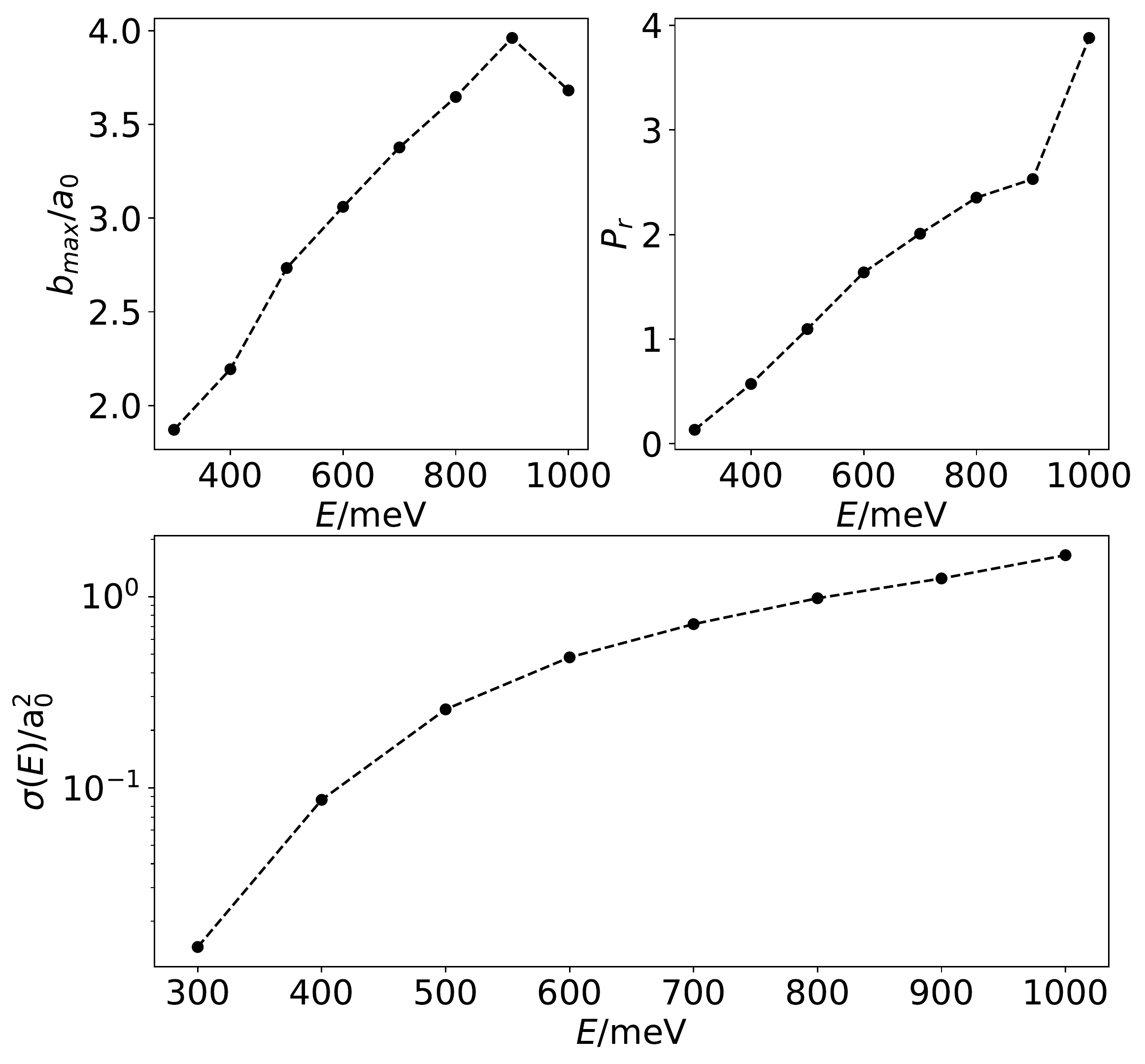}
  \caption{QCT results at fixed collisional energy for the formation of HCOOH + H. On the top
  left panel, the maximum impact parameter, on the top right panel the reaction probability and in the lower panel,
the reactive cross section.}
  \label{fig:sigma_E_HCOOH}
\end{figure}

No formation of HCOOH + H
was observed in the RPMD trajectories described below, because of the low reaction
probability of this channel and the impossibility to enrich the RPMD
statistics.
At fixed collisional energy, the only reactivity was obtained in QCT calculations above
300 meV, which is consistent with a barrier height of 276.0 meV, including ZPE.
This reaction
follows a direct mechanism, where the OH approaches the H$_2$CO
in a conformation close to the one in TS2. Indeed, the maximum impact
parameter for this process is about 3 times smaller than the one for
HCO + H$_2$O at collisional energies of 300 meV, decreasing the gap
at 1 eV to 1.5. As the collisional energy increases so does the 
reaction probability, since the system has more energy in the 
reaction coordinate. These results are summarized in figure \ref{fig:sigma_E_HCOOH}.

\begin{figure}[hbtp]
  \centering
  \includegraphics[width=0.9\linewidth]{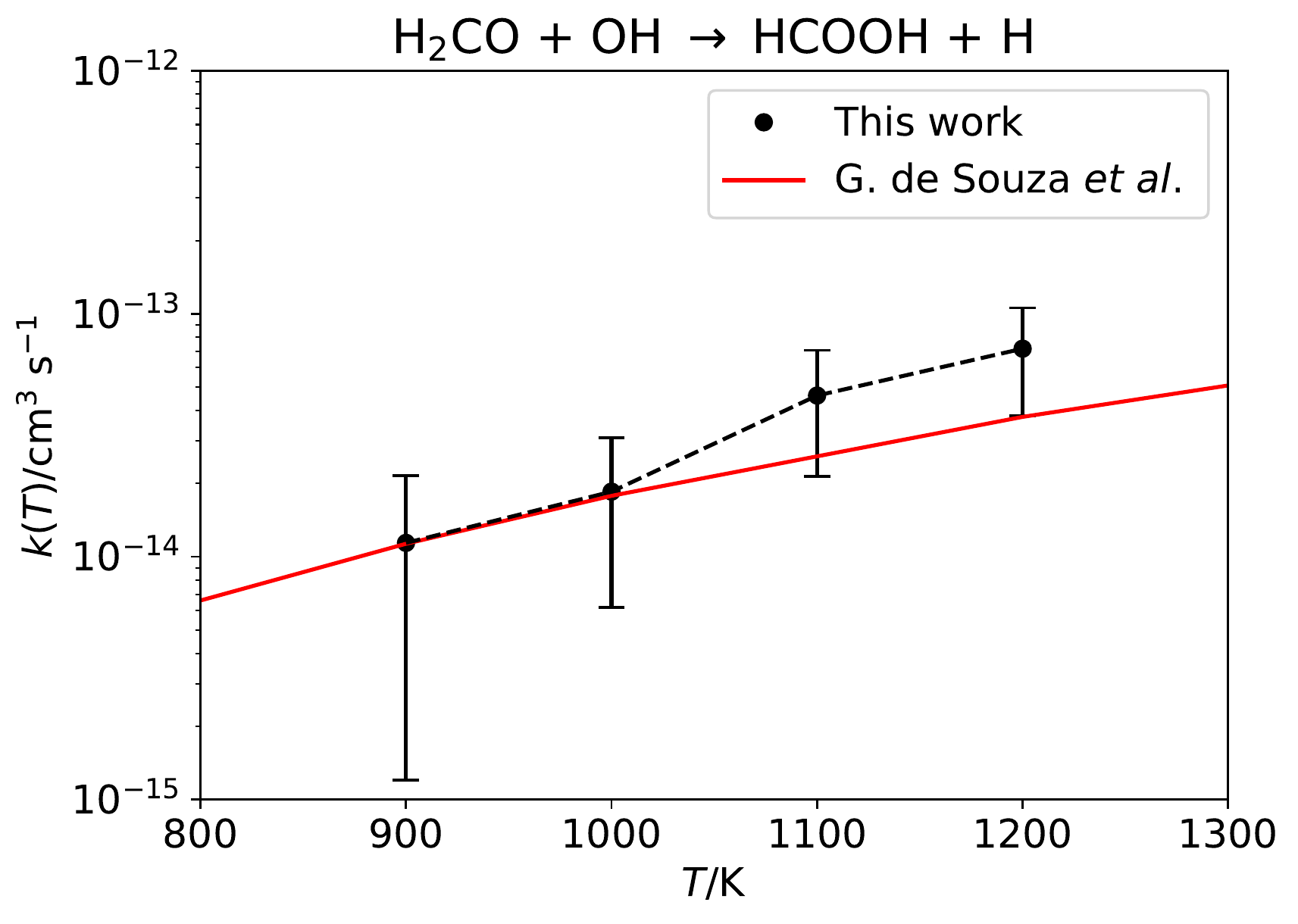}
  \caption{QCT reaction rate constant at constant temperature for HCOOH + H formation. 
The error bars are two times the standard error of the Monte Carlo
integration \cite{Karplus1965}.
	In red line
  the results obtained by G. de Souza \textit{et al.}\cite{DeSouzaMachado2020}.}
  \label{fig:rate_HCOOH}
\end{figure}

At fixed temperature, reactivity was obtained above 900 K. It is important
to remember that only reactivity from reactants in their ground vibrational
states is being taken into account. Our results are in good agreement
with the ones calculated by G. de Souza \textit{et al.} \cite{DeSouzaMachado2020}
with the canonical variational transition state method (CVTST) and are shown
in Fig. \ref{fig:rate_HCOOH}.

\subsection{RPMD results}

\begin{figure*}[hbtp]
	\begin{minipage}{0.5\linewidth}
	  \centering
	  \includegraphics[width=1.0\linewidth]{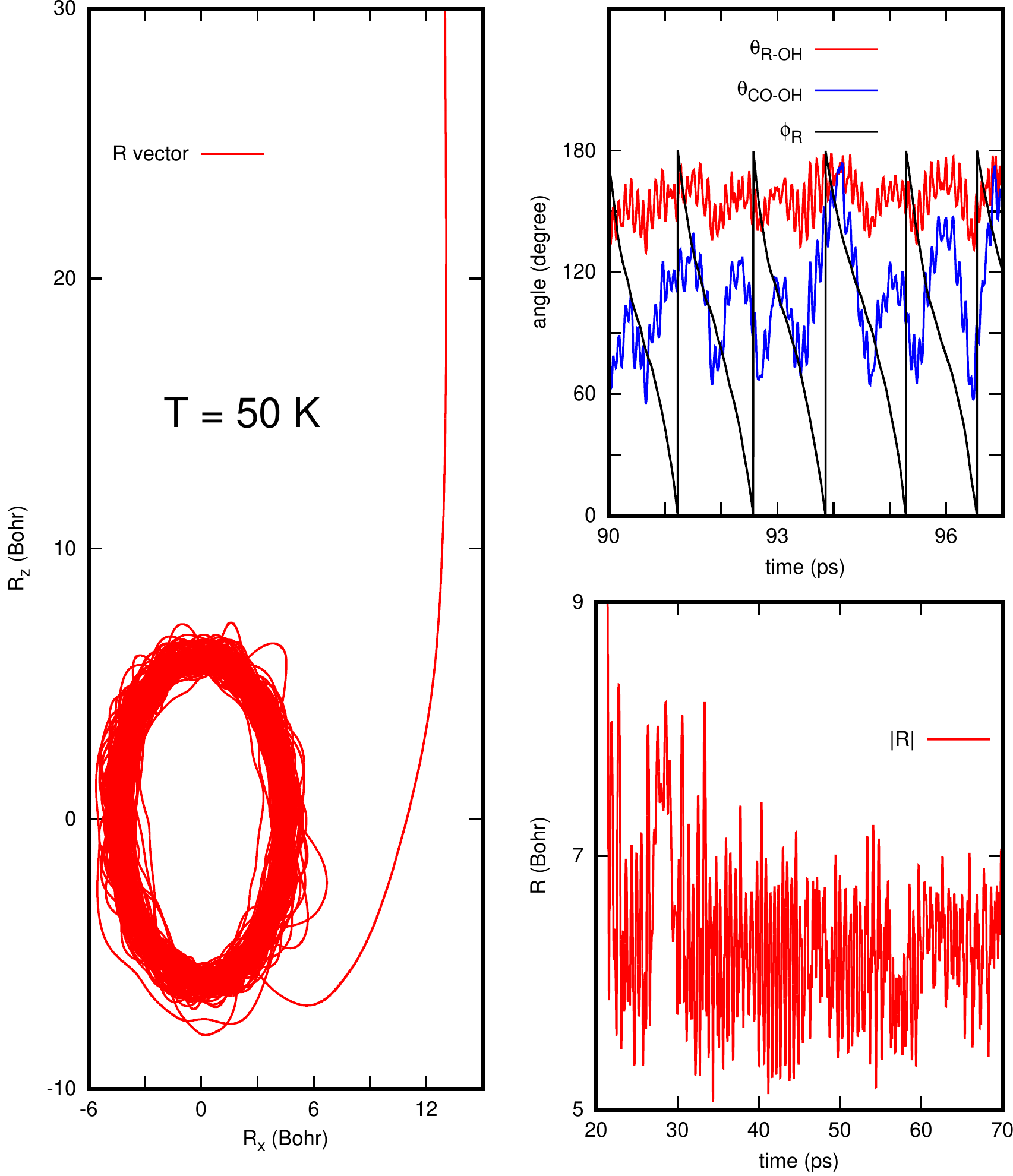}
	\end{minipage}%
	\begin{minipage}{0.5\linewidth}
	  \centering
  		\includegraphics[width=1.0\linewidth]{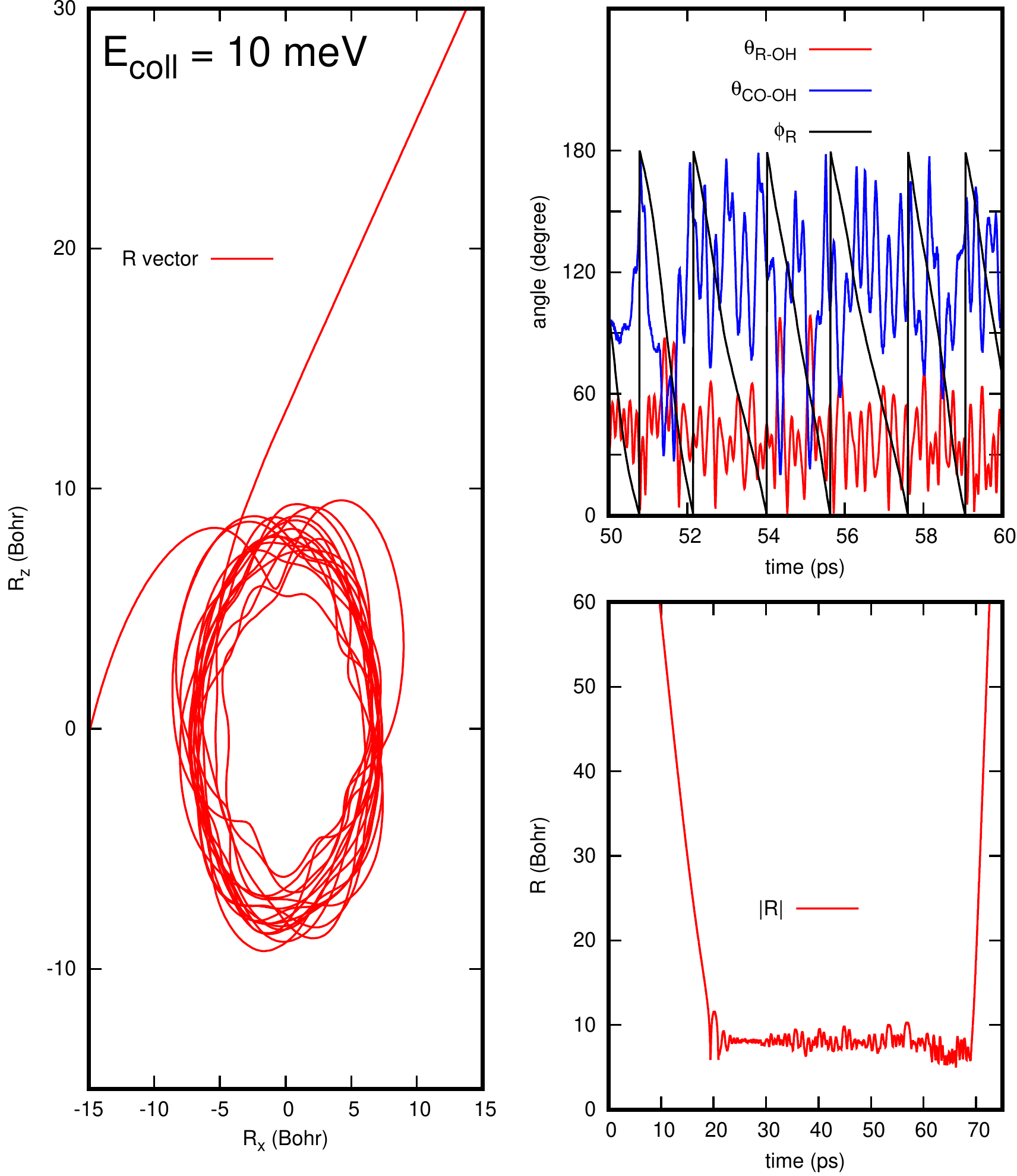}
	\end{minipage}
  \caption{Typical trapped RPMD trajectory (left 3 panels) obtained at 50 K. Right panels show a 
    typical QCT trajectory for high (red) initial impact parameter, at
    fixed collisional energy of 0.1 meV. In each case, the left panel shows the components $R_x$ and $R_z$ of the vector, {\bf R},
    connecting
    the centers of mass of the reactants  along the trajectory. The right bottom panels show the modulus of {\bf R} as a function
    of time. The right top panels show the evolution of the angles: between ${\bf R }$ and OH internuclear vector, $\theta_{R-OH}$,
   between the CO and OH internuclear vectors, $\theta_{CO-OH}$ and the azimuthal angle associated to {\bf R}, $\phi$.}
\label{fig:Rx_Rz}
\end{figure*}
Quantum effects, such as tunneling and zero-point energy, are important
at the low temperatures of interest here. Exact quantum calculations
are not feasible in this system, and  one of the most successful alternative
methods is Ring Polymer Molecular Dynamics (RPMD)
\cite{Craig-Manolopoulos:04,Craig-Manolopoulos:05,Craig-Manolopoulos:05b,%
Suleimanov-etal:11,Suleimanov-etal:16}, 
which is used in this work. RPMD is a semiclassical formalism based on Path Integral Molecular Dynamics (PIMD),
that includes quantum effects
such as ZPE\cite{RPMD:MuH} and tunneling\cite{RPMD:DMuH}.
RPMD has been successfully applied to many reactions, with barriers
or deep wells \cite{Suleimanov-etal:16,Suleimanov-etal:18}. Here we use
the dRPMD program\cite{Suleimanov-etal:18},
a direct version
of the RPMD method born as an alternative to the RPMDrate code~\cite{SULEIMANOV2013833} to deal
with reactions with no barriers. dRPMD
is parallelized to allow long propagations at low
temperatures, requiring many replicas or beads, N$_b$. The direct RPMD consists
in two steps, thermalization and real-time dynamics.

The thermalization is a constrained PIMD simulation, using
the Andersen thermostat \cite{Andersen1980} and freezing the
distance between the two reactants at 120 $a_0$. This propagation
is performed during 10$^4$-10$^5$ steps, depending on the temperature,
to warrant convergence of
the initial ZPE of each fragment. At the end of the thermalization 
the relative position and velocities of the two reactants are reoriented,
imposing a maximum initial impact parameter, similar to QCT calculations.

In the second step, the constrain and the thermostat
are removed, leading to the classical evolution of a system
consisting on $N_{atoms}\times N_{b}$ particles, which are propagated
using a second order simplectic operator.
In previous RPMD calculations \cite{delMazo-Sevillano-etal:19},
the simplectic propagation is separated
in two steps: first the free polymer and second the real potential
terms. The free polymer
propagation was done using a Fourier transform (FT) \cite{SULEIMANOV2013833}, and, being
nearly exact,  allows a large time step.
In the previous RPMD calculations \cite{delMazo-Sevillano-etal:19,Naumkin-etal:19}
it was also found that, for temperatures below 100 K, many trajectories
were trapped in the complex well in the reactants channel. Those trapped
trajectories lived for more than 500 ns, and were too long to be finished.
So long-lived complexes,
as compared to those obtained in QCT calculations (of several ps), could
be explained by the reduction of the available reactants phase space
produced in RPMD, in which ZPE is taken into account. However, it is
also known that RPMD may show spurious resonances
\cite{Witt-etal:09,Rossi-etal:14,Jang-etal:14,Korol-etal:19}, which
can be attributed to polymer normal mode excitations probably accessed
when the frequencies of these modes are close to the  frequencies
associated to the ``physical'' system. These spurious resonances
seem to be enhanced when the free ring polymer is propagated as
a separate entity and several alternatives have been proposed \cite{Jang-etal:14,Korol-etal:19}.

The choice made in this work is not to propagate the free ring polymer. Instead,
we separate the ring polymer Hamiltonian in physical kinetic energy and total
potential, containing the PES among the $N_{atoms}$ atoms  and the harmonic oscillators among
the beads of the same atoms. This implies to reduce the time step to approximately
$\Delta t /\sqrt{N_{b}} $, with $\Delta t$ the time step used in the FT propagation.
This reduction of the time step increases the computational time, but
when using parallel computers the present factorization
reduces the communication to only neighbor processors,
thus allowing a higher speed up of the parallelization.

The parameters used for the calculation of the RPMD rate constants for different temperatures
are listed in table~\ref{tab:RPMD_parameters}. At the end of each RPMD trajectory, the
product channel is determined by analyzing the centroid, similarly
to what is done in QCT calculations. What is different in RPMD calculations, is that
there is an increasing  probability of trapped trajectories as temperature decreases below 300 K.
Those trapped trajectories are formed following an orbit similar to the QCT orbit of Fig.~\ref{fig:Rx_Rz}
as those shown in previous RPMD calculations in this system\cite{delMazo-Sevillano-etal:19,Naumkin-etal:19}:
the long range interaction deviates the initial trajectories for rather long impact parameters,
increasing the end-over-end angular momentum and the rotation of each of the
two reactants. These trajectories get then trapped orbiting continuously keeping
the relative orientation of the two reactants fixed, close to the geometry of reactant well, RC1.
What is different from QCT calculations, is that these trapped orbits live very long times,
longer than $t_{max}$ = 10 ns in this case, when they are stopped. Thus we give separate probabilities
for direct (reactive RPMD trajectories) and trapped trajectories in table~\ref{tab:RPMD_parameters}.
The rates for direct  reaction and trapping processes are then evaluated through
equation \eqref{eqn:k_T}, as listed in table~\ref{tab:RPMD_parameters}

\begin{table*}
  \caption{\label{tab:RPMD_parameters}Parameters of the direct RPMD and rate constants.}
\begin{ruledtabular}
  \begin{tabular}{c c c c c c c c c c}
	$T$/K   & $N_{b}$ & $N_{total}$ & $t_{max}$/ns & $b_{max}^{dir}/a_0$ & $b_{max}^{trap}/a_0$ & $P_{dir}$ & $P _{trap}$  & $k_{dir}$ / cm$^3$ s$^{-1}$ & $k_{trap}$/ cm$^3$ s$^{-1}$ \\
\hline
1200 & 144  & 10000 & $1.0$ & $8.44 $ & $0.00 $ & $0.025$ & $0.00$ & $1.31\times 10^{-11}$ & 0$.00$ \\
1000 & 48   & 2454  & $0.1$ & $5.82 $ & $0.00 $ & $0.048$ & $0.00$ & $1.10\times 10^{-11}$ & 0$.00$ \\
800  & 144  & 3749  & $0.1$ & $7.93 $ & $0.00 $ & $0.019$ & $0.00$ & $7.21\times 10^{-12}$ & 0$.00$ \\
600  & 144  & 4649  & $0.1$ & $6.76 $ & $0.00 $ & $0.026$ & $0.00$ & $6.67\times 10^{-12}$ & 0$.00$ \\
400  & 96   & 6695  & $0.1$ & $6.93 $ & $0.00 $ & $0.016$ & $0.00$ & $3.82\times 10^{-12}$ & 0$.00$ \\
300  & 240  & 30000 & $1.0$ & $11.93$ & $7.86 $ & $0.003$ & $2.59\times 10^{-4}$ & $2.18\times 10^{-12}$ & 7$.11\times 10^{-14}$ \\
200  & 384  & 20000 & $1.0$ & $11.78$ & $13.82$ & $0.005$ & $2.01\times 10^{-3}$ & $2.58\times 10^{-12}$ & 1$.54\times 10^{-12}$ \\
100  & 768  & 1000  & $1.0$ & $12.95$ & $16.60$ & $0.003$ & $1.94\times 10^{-1}$ & $1.92\times 10^{-12}$ & 1$.83\times 10^{-10}$ \\
50   & 1536 & 400   & $1.0$ & $13.39$ & $18.96$ & $0.015$ & $5.89\times 10^{-1}$ & $7.15\times 10^{-12}$ & 5$.72\times 10^{-10}$ \\
20   & 3072 & 143   & $1.0$ & $14.53$ & $21.43$ & $0.048$ & $6.40\times 10^{-1}$ & $1.75\times 10^{-11}$ & 5$.11\times 10^{-10}$ \\
  \end{tabular}
\end{ruledtabular}

\end{table*}

\begin{figure}[hbtp]
  \centering
  \includegraphics[width=.9\linewidth]{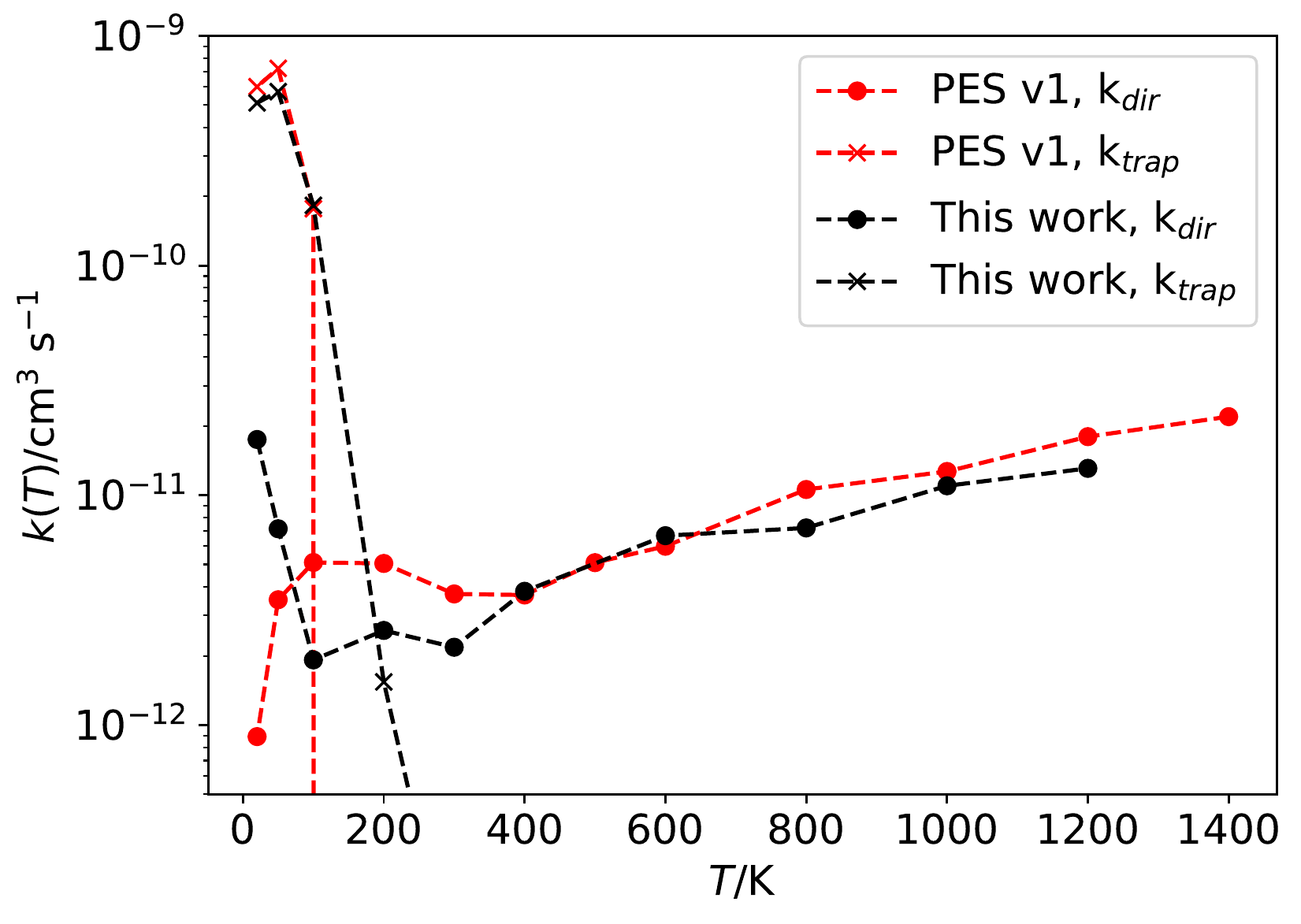}
  \caption{Reaction and trapping rates for the present PES, compared
  with the ones from out previous fit.}
  \label{fig:reac_trap_k}
\end{figure}

The trapping mechanism becomes significant at temperatures below
200 K, where trajectories finished after 10 ns without dissociation or
reaction. This results perfectly mimic the ones from our previous
work \cite{delMazo-Sevillano-etal:19} as shown in figure \ref{fig:reac_trap_k}.
As compared to our previous PES, the trapping mechanism begins
at temperatures slightly higher, around 300 K. With respect to the 
direct mechanism rate, the difference found in the QCT study seems to
magnify, what on top of what has already been said in the previous
section, it is worth noting that some of the errors may emerge from
a poorer statistic, due to the huge computational cost of RPMD
trajectories at these low temperatures.
A typical trapped RPMD trajectory is  shown in the right
panels of Fig.~\ref{fig:Rx_Rz}, compared with a direct and complex-forming
QCT trajectories shown in the right panels of  Fig.~\ref{fig:Rx_Rz}.
The first difference in both trajectories is that RPMD trajectory keeps trapped
for much longer times than QCT trajectory, that actually manages to break 
the reactive complex in the propagation time. Once the two reactants collide,
$\phi_R$ oscillates quasi periodically in both cases, but
the angle $\theta_{R-OH}$  vector keeps almost constant. This behaviour is endorsed by the
$\theta_{CO-OH}$ angle, and the only appreciable difference is that QCT
amplitudes have larger variations.

With respect to our previous work  \cite{delMazo-Sevillano-etal:19}, not
only the partition of the RP Hamiltonian has been varied, but also the
time step, $\Delta t$, has been reduced by a factor of 10 and the
number of beads used in the low temperature calculations has 
been considerably increased, in order to check whether or not   long-lived complex
lifetimes are  artificially due to
poor convergence. The computational cost of this PES has been considerably
reduced, what has enabled to perform this kind of 
study. However, the formation of extremely long trapped trajectories
persist here, as can be seen in figure \ref{fig:reac_trap_k}.
Therefore, we discard the possibility of this results being
a lack of convergence of the calculations.
Similar behaviour was also found for other reactions at low temperature,
OH + CH$_3$OH\cite{delMazo-Sevillano-etal:19},
D + H$_3^+$\cite{Bulut-etal:19} and H$_2$ + H$^+_3$\cite{Suleimanov-etal:18}.
We therefore conclude that the trapping is a rather general outcome of
RPMD calculations at low temperature.

\begin{figure}[hbtp]
  \centering
  \includegraphics[width=.9\linewidth]{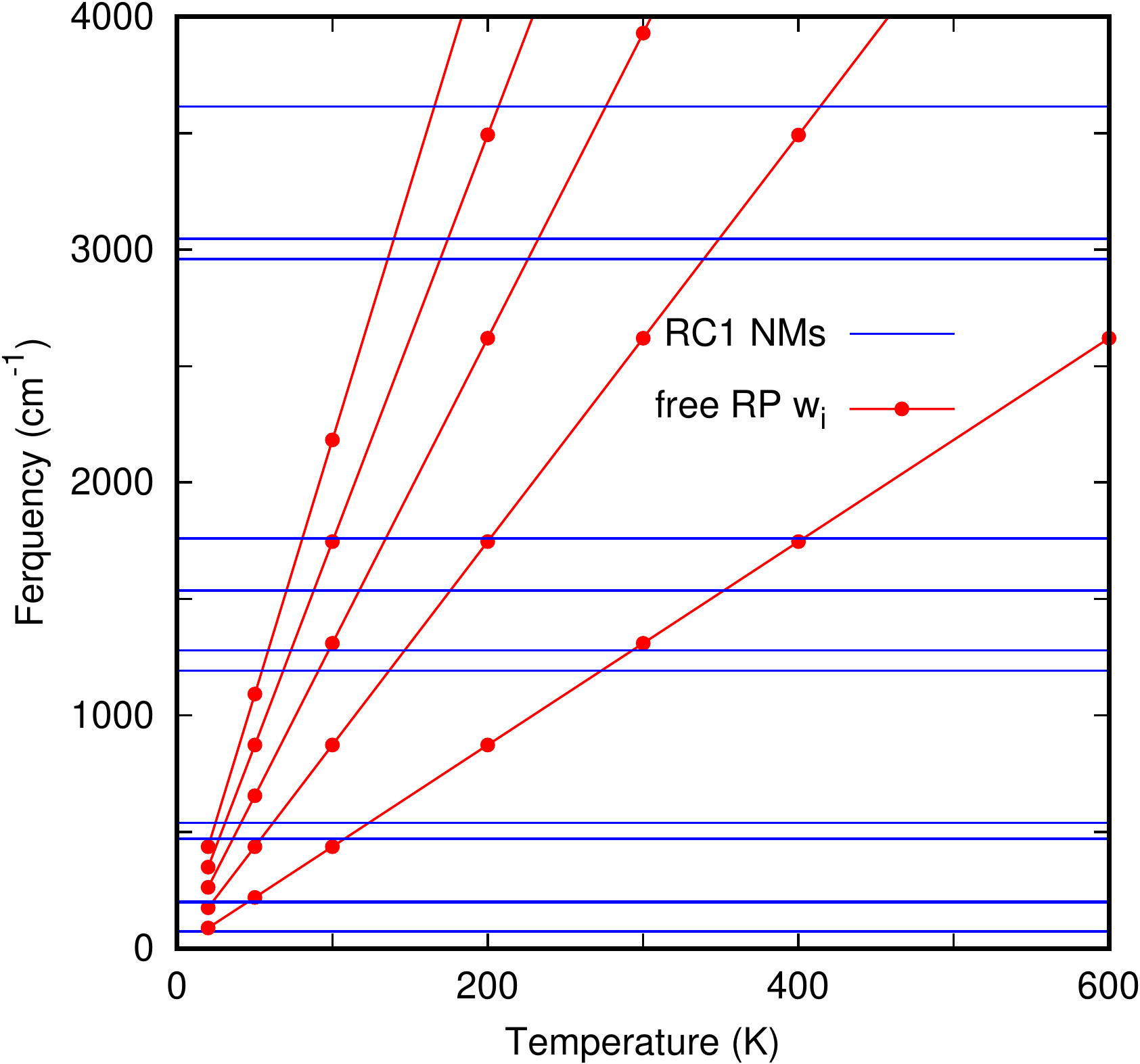}
  \caption{RC1 complex normal mode frequencies (blue) and  first five $\omega_i$ free ring-polymer
    frequencies versus temperature.}
  \label{fig:rpNM}
\end{figure}
An increase of complex lifetimes  when including quantum effects is expected
and can be explained
according to the statistical theory RRKM\cite{doi:10.1063/1.1700422, doi:10.1063/1.1700423} 
by the descent of accesible dissociative states in the RPMD calculations with
respect to the QCT ones, because ZPE is included. However, the impossibility to
end the trapped trajectories, dominant at low temperatures, seems to indicate
that RPMD method fails to reproduce the real lifetime of the reaction complexes.
This is attributed to the appearance of spurious resonances as those reported
previously \cite{Witt-etal:09,Rossi-etal:14},  that appear when the frequency of
the complexes are of the order of those of the ring polymer modes. The normal
modes of the free ring polymers \cite{Richardson-Althorpe:09} are
given by $\omega_k = 2N_{b} k_B T /\hbar \sin(\vert k\vert \pi/N_b)$, with $k_B$ being the Boltzmann constant and
$-N_b/2 \leq k \leq N_b/2$. The lowest frequency mode, $k=1$,  for large $N_b$
can be approximated by  $\omega_1\approx 2 \pi T k_B /\hbar$, and when this quantity
is of the order of the lower frequency of the reaction complex, the energy
flow between the ``physical'' and ``free ring-polymer'' modes may be enhanced,
giving rise to spurious resonances, in which the energy is stored in the
free ring-polymer normal modes.
The lower $\omega_i$  are plotted as a function of temperature and compared with the
physical frequencies of the H$_2$CO-complex in Fig.~\ref{fig:rpNM}. Clearly
at $\approx$ 100 K, the first RP normal mode is of the order of the lower RC1 normal modes.
As temperature decreases, there are Ring-polymer modes in near resonance with
several physical modes of the RC1 complex, favoring the energy transfer.
 Since the density of the RP modes is large, the energy is stored there
for very long times, giving rise to artificial trapping or spurious resonances.

\begin{figure}[hbtp]
  \centering
  \includegraphics[width=.9\linewidth]{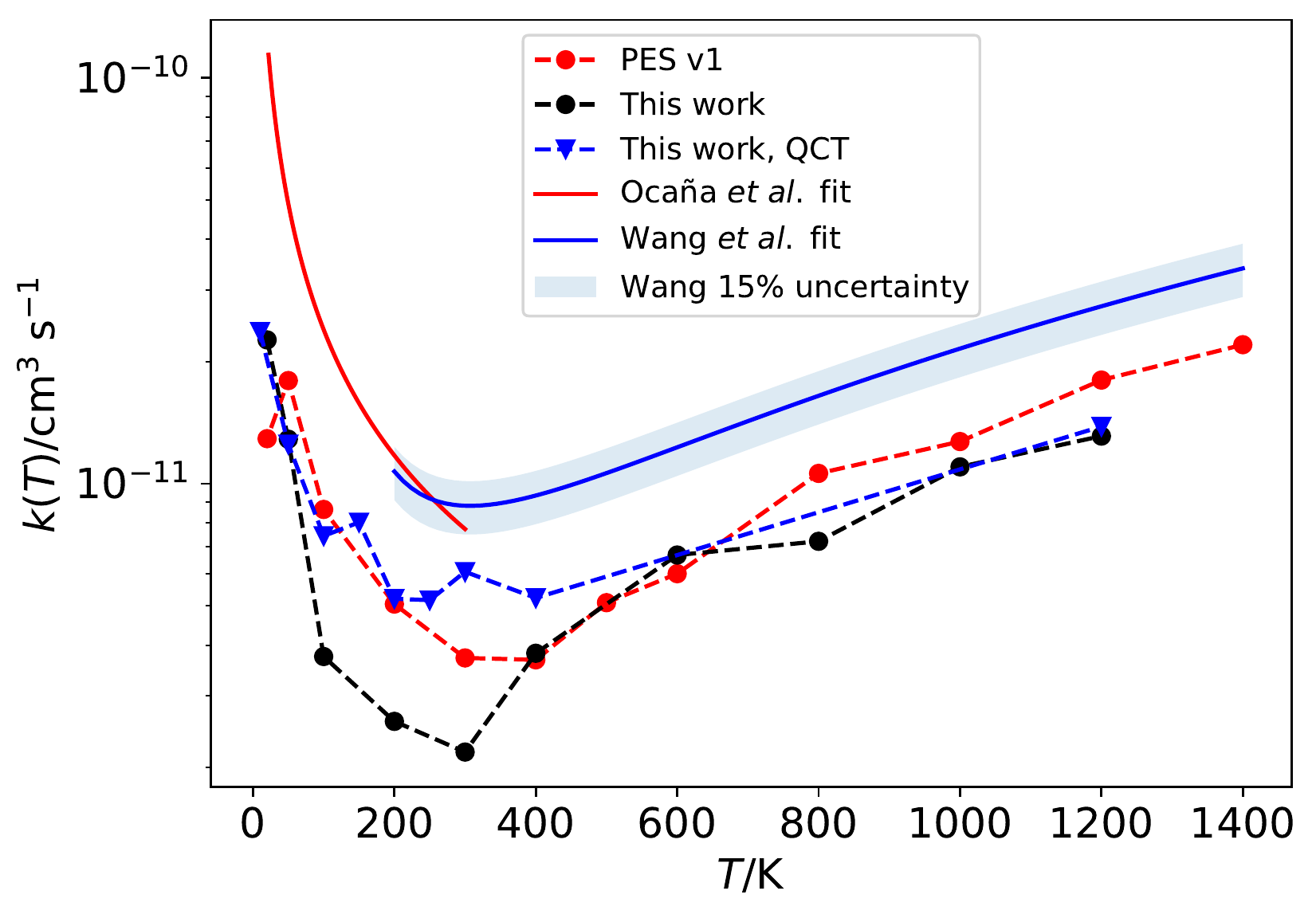}
  \caption{
	  RPMD reactive rate constant for HCO + H$_2$O formation.
    Black points represent the RPMD results obtained in this work. 
	Blue triangles represent the QCT results obtained in this work.
Red points are the results from our previous PES \cite{Zanchet-etal:18}. 
Solid lines show the recommended fit from Ocaña \textit{et al.} \cite{Ocana-etal:17}
in red and from Wang \textit{et al.} \cite{Wang2015} in blue.
}
  \label{fig:RPMD_rate}
\end{figure}

Several solutions have been proposed to solve the problem of spurious resonances
in RPMD calculations of spectra  \cite{Witt-etal:09,Rossi-etal:14}. Among them,
the inclusion of a thermostat to the internal modes
of the ring polymer during the dynamics \cite{Rossi-etal:14}
is considered here that may affect the reaction dynamics, because the extra
energy of the beads is expected to flow to the physical bonds, producing their
fragmentation. A detailed analysis needs to be done before applying it
 to reactive collisions at low temperatures. Instead,
in order to get the total reaction rate constant we  have to
look for an alternative method to consider the fragmentation ratio of the trapped trajectories:
back to reactants (redissociation) and tunneling through reaction barriers to products (tunnel).
Under this assumption, the total reaction rate becomes: 
\begin{eqnarray}
  k(T) &=& k_{dir}(T) + k_{CF}(T)  \\
  k_{CF}(T) &=& k_{trap}(T) \frac{k_{tunnel}(T)}{k_{tunnel}(T) + k_{rediss}(T)},
\end{eqnarray}
where $k_{dir}$ is the rate constant for the direct mechanism and $k_{CF}$ 
is the product of the trapping rate constant and
the ratio of tunneling trajectories towards products. This approach was suggested
before\cite{delMazo-Sevillano-etal:19,Naumkin-etal:19}, and the ratios were
obtained either from TST or QCT methods. Here we adopted the QCT reaction probability
as
\begin{equation}
  \frac{k_{tunnel}(T)}{k_{tunnel}(T) + k_{rediss}(T)} \approx P_r^{QCT}(T),
\end{equation}
that is, $\approx 1\%$ for collision 
energies below 1 meV. This classical estimate is justified by the
small height of the
TS, that can even be reduced due to anharmonic effects. The resulting
RPMD reaction
rate constants are shown in figure \ref{fig:RPMD_rate} compared with previous results.
The behaviour of both results are similar and always below the experimental
values.

  Black points and blue triangles represent the RPMD and QCT results obtained in this work, respectively. 
Red points are the RPMD results from our previous PES \cite{Zanchet-etal:18}. 
Solid lines show the recommended fit from Ocaña \textit{et al.} \cite{Ocana-etal:17}
in red and from Wang \textit{et al.} \cite{Wang2015} in blue.

The difference between simulated and experimental rate constants for T$>$400 K
is attributed to the appearance of other mechanisms, such as non-adiabatic
contributions of excited electronic state, and/or a lack of accuracy to
describe other channels such as H + HCOOH  product channel. In this work
we focus on the lower temperature behaviour. Here the problem is to determine
the tunneling/redissociation fraction of the trapped RPMD trajectories. Such
study is fundamental for astrochemistry, since it is necessary to use the pure
zero-pressure rate constant for the formation of OM in cold environments.

Also, it is important to determine the life-time of the reaction complexes to
establish the role of complexes (or pressure) in the measured rate constants
in CRESU experiments. Work in this direction is now on the way, following the
preliminary arguments already published\cite{Naumkin-etal:19}.

\section{Conclusions}

In this work a new full dimensional potential energy surface has been developed for
the title reaction based on artificial neural networks. The long-range behaviour
of the ANN has been analyzed in detail. A source of spurious interactions has been 
identified due to the non linearity of the transfer function in the ANN, that mixes the
internal degrees of freedom of each of the polyatomic fragments.
These spurious interactions change the energy of the fragments at very
long distances introducing artifacts in the dynamics at low energies, thus
becoming inappropriate to study reactivity at low temperatures.

To solve this problem, a new factorization of the potential is proposed consisting
in two terms, a diabatic matrix and a full dimensional term, $V^{MB}$, expanded using neural networks.
In the diabatic matrix, the diagonal  elements describe each rearrangement channel, including
the potential of each independent fragment plus their interaction among them, with
the long range interaction properly set. Thus, $V^{MB}$ fits the difference and tends
to zero at intermediate distance. This term is further multiplied by a switching
function to fully remove the spurious long range interaction introduced by the artificial neural network
function. The present potential is more accurate than the previous one\cite{Zanchet-etal:18}
and also incorporates the channel towards HCOOH + H product channel.

This PES has been used to calculate the reaction rate constants using QCT and RPMD methods.
It is found that the HCOOH + H products present a near negligible contribution
at the energy range considered. The HCO + H$_2$O reaction rate constant presents a non-Arrhenius V-shape as a 
function of temperature, dominated by a direct mechanism at high temperatures and a complex-forming
at low energies, as it was also found in the previous PES\cite{delMazo-Sevillano-etal:19} and
in qualitative agreement with the available experimental data \cite{Ocana-etal:17}.

In spite of the changes done in the calculations, the RPMD results show an increasing trapping probability
as temperature decreases, as reported before\cite{delMazo-Sevillano-etal:19}. This is attributed
to the presence of spurious resonances occurring when the free ring-polymer normal mode frequencies
enter in near resonance with the low intermolecular normal modes frequencies. It is crucial to
determine the life-time of these complexes and the fragmentation ratio in order to 
properly determine the zero-pressure reaction rate constant, needed in astrophysical models, and
to determine the role of complexes in the measurements made in laval expansions. Some work in these directions
are now-a-days under way.

\section{Supplementary Material}

See supplementary material for the multi-reference calculations
of ground and first electronic states along the MEP, the diabatic matrix, $V^{diab}$
construction, the hyper-parameters used in the NN fits and the normal mode frequencies in the transition
states.\\

A Fortran 90 implementation of this potential energy surface is provided in the Supplementary Information.

\section{Acknowledgements}
The research leading to these results has received funding from
MICIU (Spain) under grant FIS2017-83473-C2.
We also acknowledge computing time at Finisterre (CESGA) and Marenostrum (BSC)
under RES computational grants ACCT-2019-3-0004 and  AECT-2020-1-0003.

\section{Data availability}
The data that support the findings of this study are available from the corresponding author
upon reasonable request.


%
\end{document}